\documentclass[12pt,english]{article}
\usepackage[T1]{fontenc}
\usepackage[latin9]{inputenc}
\usepackage{geometry}
\usepackage{xcolor}
\usepackage{float,cite}
\usepackage{booktabs}
\usepackage{amstext}
\usepackage{graphicx}
\usepackage{authblk}

\makeatletter

\providecommand{\tabularnewline}{\\}
\newcommand{\lyxdot}{.}

\makeatother

\usepackage{babel}
\begin{document}
\title{A short review on quantum identity authentication protocols: How would
Bob know that he is talking with Alice?}
\author[1,$\$$]{Arindam Dutta}
\author[1,*]{Anirban Pathak}
\affil[1]{Jaypee Institute of Information Technology, A 10, Sector 62, Noida, UP-201309, India}
\affil[$\$$]{arindamsalt@gmail.com}
\affil[*]{anirban.pathak@gmail.com}
\date{}
\maketitle

\begin{abstract}
Secure communication has achieved a new dimension with the advent
of the schemes of quantum key distribution (QKD) as in contrast to
classical cryptography, quantum cryptography can provide unconditional
security. However, a successful implementation of a scheme of QKD
requires identity authentication as a prerequisite. A security loophole
in the identity authentication scheme may lead to the vulnerability
of the entire secure communication scheme. Consequently, identity
authentication is extremely important and in the last three decades
several schemes for identity authentication, using quantum resources
have been proposed. The chronological development of these protocols,
which are now referred to as quantum identity authentication (QIA)
protocols, are briefly reviewed here with specific attention to the
causal connection involved in their development. The existing protocols
are classified on the basis of the required quantum resources and
their relative merits and demerits are analyzed. Further, in the process
of the classification of the protocols for QIA, it's observed that
the existing protocols can also be classified in a few groups based
on the {[}inherent computational tasks used to design the protocols.
Realization of these symmetries has led to the possibility of designing
a set of new protocols for quantum identity authentication, which
are based on the existing schemes of the secure computational and
communication tasks. The security of such protocols is also critically
analyzed.
\end{abstract}

\section{Introduction\label{sec:Introduction}}

Authentication or identity authentication is a systematic procedure
of validating the identity of the legitimate users and/or components/devices.
This process helps us to circumvent various kinds of attacks on the
schemes for secure computation and communication. The relevance of
the schemes of identity authentication has been considerably increased
in the recent past with the enhanced use of online banking, e-commerce,
internet of things (IoT), online voting, etc. All these applications,
and many others essentially require identity authentication. For example,
IoT primarily requires authentication of the devices or components,
whereas online voting requires authentication of the users (voters).
From the perspective of cryptography, there is not much difference
between a user and a component, and in what follows, we will treat
them as equivalent. 

In a two party scenario, authentication may be visualized as a procedure
that allows a legitimate user (sender) Alice to transmit a message
$X$ (equivalently a key in case of key distribution schemes) to a
second user Bob (receiver) in such a way that Bob can be assured that
the data was not corrupted during transmission through the channel
\cite{DL99}. In other words, it can be viewed as a procedure that
certifies the identity of the creator of the message $X$ and establishes
the integrity of the message received by Bob. This is closely related
to digital signature, which would allow a third party (Charlie) to
verify at a later time that Bob has not modified the message $X$
sent to him by Alice \cite{DL99}. Authentication and digital signature
are closely related but different tasks, and in the present review,
we will restrict ourselves to the schemes for authentication only.
More precisely, this review will be restricted to the schemes of authentication,
which uses quantum resources, i.e., schemes for quantum identity authentication
(QIA). To appreciate the relevance and importance of QIA schemes,
we first need to briefly discuss the development of quantum cryptography
which had historically transformed the notion of security.

The first ever protocol of quantum key distribution was proposed by
Bennett and Brassard in 1984 \cite{BB_1984} which is now known as
BB84 protocol\footnote{Of course, the origin of quantum cryptography owes a lot to the pioneering
work of Wiesner \cite{W83}. For a beautiful description of the related
history see \cite{B05}.}. The claim that this protocol can provide unconditional security
which is a desired facet not achievable by classical schemes of key
distribution where security arises from the complexity of a mathematical
task drew considerable attention of the cryptographic community. This
led to a set of interesting protocols of quantum key distribution
(QKD) including E91 \cite{ekert1991quantum} and B92 \cite{bennett1992quantum}
protocols for QKD. The need for the identity authentication of the
legitimate users were recognized in these works as we can see that
even in the classic BB84 paper Bennett and Brassard wrote, ``The
need for the public (non-quantum) channel in this scheme to be immune
to active eavesdropping can be relaxed if the Alice and Bob have agreed
beforehand on a small secret key, which they use to create Wegman-Carter
authentication tags \cite{WC81} for their messages over the public
channel''. However, as no quantum identity authentication protocol
was available prior to 1995, classical schemes for identity authentication,
like Wegman-Carter scheme was intrinsically considered in the early
schemes for QKD, although it was not always explicitly spelled out.
In all such schemes and also the schemes for authentication developed
later, ``pre-shared small key'' mentioned above plays a crucial
and essential role\footnote{In fact, all the protocols of QKD can be viewed as key amplification
protocols as every QKD protocol starts with a pre-shared small key
used for authentication and creates a longer key in a particular session,
and subsequently uses a small part of the longer key as the pre-shared
small key in the next session.}. This point will be further clarified in the later part of this paper.
The use of Wegman-Carter scheme or a similar scheme in the early protocols
of QKD was a natural consequence of the fact that no real channel
can be strictly considered as faithful. Specifically, an eavesdropper
(Eve) can replace a channel between the sender (Alice) and the receiver
(Bob) by two channels, one from Alice to Eve and other from Eve to
Bob. In such a scenario that can separate Alice and Bob, Eve will
able to create a key with Alice and another with Bob. The need to
circumvent such a scenario makes it essential to use a protocol for
identity authentication.

\begin{figure}

\begin{centering}
\includegraphics[scale=0.3]{ 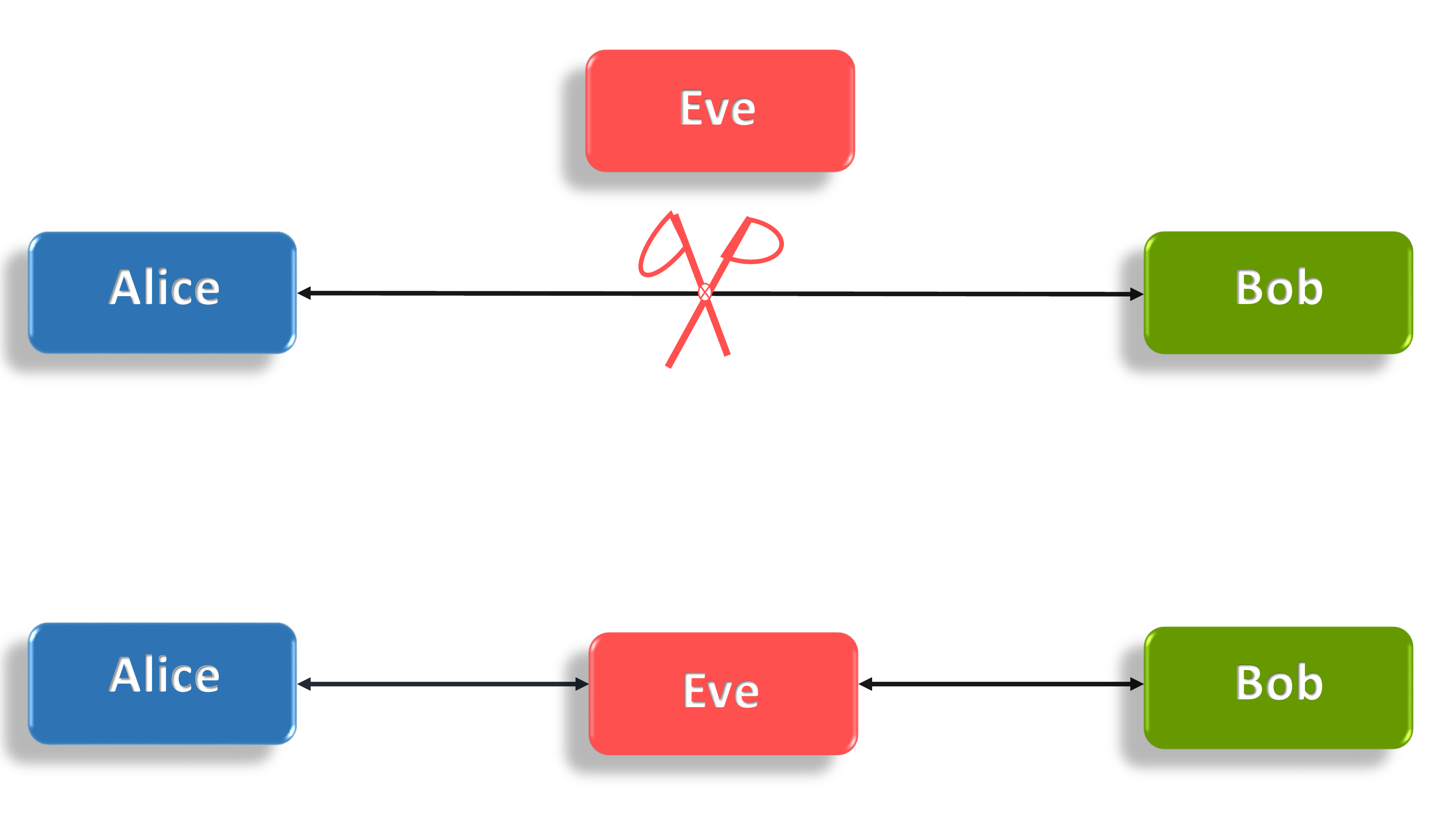}\caption{\label{fig:1 Eve-is-trying} (Color online) Eve may try to impersonate
Alice (Bob) in front of Bob (Alice), by replacing the channel between
Alice and Bob with two channels- one from Alice to Eve and another
from Eve to Bob.}
\par\end{centering}
\end{figure}
Now, as the security of the classical authentication schemes are also
based on the complexity of the computational task(s) involved, authentication
using classical resources cannot be done in an unconditionally secure
manner, and consequently it may lead to a security loop-hole(s) into
the so called unconditionally secure quantum key distribution schemes.
Thus, to obtain a really unconditional secure scheme of quantum key
distribution, one would require an unconditional security for the
authentication scheme, too,\textcolor{brown}{{} }and that in turn would
need to use quantum resources. Such a scheme of identity authentication,
which uses quantum resources is generally referred to as a quantum
identity authentication (QIA) scheme. First such scheme was proposed
by $\text{{\rm Cr�peau}}$ et al. \cite{CS_1995} in 1995, and it
was followed by a large number of schemes. For example, before the
flooding of schemes for QIA, in 1998, Zeng et al. proposed a QKD protocol
\cite{ZW_1998} that allowed the simultaneous distribution of the
key and the verification of the communicator's identity and almost
immediately after that in 1999, Du�ek et al. \cite{DHHM_1999} proposed
two protocols for QIA. The protocols were hybrid (in the strict sense,
this is apparently the case with all the schemes proposed till date)
in the sense that they were proposed using an amalgamation of classical
authentication schemes and protocols for QKD with a specific stress
on one time pad. The sequence continued and a large number of protocols
having individual advantages and disadvantages appeared \cite{LB_2004,WZT_2006,ZLG_2000,ZZZX_2006,ZCSL_2020,KHHYHM_2018,CXZY_2014,T.Mihara_2002}.
Considering the importance of secure quantum computation and communication
protocols, and the crucial role of authentication in those protocols,
here we aim to systematically review the protocols for QIA.

The rest of the paper is organized as follows. In Section \ref{sec:A-short-chronological history},
we have provided a short chronological history of the development
of the schemes for QIA with specific attention to some schemes which
are distinct in their characteristics. In Section \ref{sec:resource-based-Classification}
and Section \ref{sec:task-based-Classification}, we have classified
the existing schemes for QIA based on the resources and computational/communication
tasks used to implement a scheme, respectively. Subsequently, a set
of new protocols are proposed in Section \ref{sec:New-protocols};
and in Section \ref{sec:Comparison}, the proposed protocols and the
existing protocols are compared. The limitations of the particular
classes of protocols are also highlighted. The security of the proposed
protocols are analyzed in Section \ref{sec:Security-analysis}. Finally,
the paper is concluded in Section \ref{sec:Conclusions}. 

\section{A short chronological history of the protocols for quantum identity
authentication \label{sec:A-short-chronological history}}

We have already mentioned in Section \ref{sec:Introduction}, that
classical authentication scheme, like Wegman-Carter scheme, was used
in the early protocols for QKD. Such message authentication schemes
can lucidly be described as schemes composed of two phases. In the
first phase,\textcolor{green}{{} }functions which produce identity authentication
are cleverly utilized and in the second phase, the authenticity of
a message is verified. Classically functions, like message encryption,
message authentication code (MAC) and Hash functions are commonly
used for authentication. Although, it's not our focus to discuss classical
authentication schemes, we can briefly and lucidly mention some relevant
ideas before we provide a chronological history of QIA. Such a short
discussion on classical tricks is justified, as hybrid and completely
classical schemes are still used in various schemes for secure quantum
communication and computation. Further, some of the classical tricks
of authentication can be easily extended to design the schemes for
QIA. Let's first establish the point that classical tricks can be
extended to designing of QIA through an example. Consider a pre-shared
key $x_{1},x_{2},\cdots,x_{n}$ between Alice and Bob, where $x_{i}\in\{0,1\},$
now Alice can use half of this as a message (say all the odd bits)
and other half as key (say all the even bits) and create a new sequence
of length $\left\lceil \frac{n}{2}\right\rceil $ such that $i$th
bit of the new sequence is $x_{2i-1}\oplus x_{2i}$. This can be viewed
as a Tag or equivalently a crypt. Alice can send it in a conventional
manner and Bob can use the same key $\{x_{2i}\}$ to decode it. Capability
of communicating a message ensures the capability of communicating
a Tag. So every scheme of quantum secure direct communication (QSDC)
\cite{YSP14}, quantum dialogue (QD) \cite{SKB+13}, deterministic
secure quantum communication (DSQC) \cite{SP14} can perform this
task. In other words, the above classical trick can be extended easily
to form simple minded schemes for QIA using different protocols for
QSDC, DSQC and QD. Further, we may note that in the MAC technique,
both parties share a common secret key $K$. When Alice wants to send
a message to Bob, she calculates the MAC as a function of the message
and the key, and sends the message and MAC to Bob. Subsequently, Bob
calculates the MAC information in the same manner and compares that
with the received MAC. If it matches, the receiver confirms the authenticity
and also the integrity of the message. A hash function is a variation
of MAC, which takes variable size of messages as input and produces
a fixed-size of output referred to as hash code. Hash function has
a one-way property and consequently, it's useful to produce a fingerprint
of a message.

Now, it's important to note that no-cloning theorem which states that
an unknown quantum state cannot be copied with unit fidelity and the
collapse on measurement property of quantum states plays a crucial
role in establishing advantages of the schemes for QIA over their
classical counterparts. To visualize this let us consider a situation,
where Eve trespasses into Alice's house in her absence and makes a
copy of her authentication key without leaving any trace. In a classical
scenario, this will allow Eve to pose as Alice and communicate with
Bob without being detected. However, due to no-cloning theorem such
a situation would not arise in the quantum world, if the authentication
key is encoded and stored in a quantum state. Further, due to collapse
on measurement principle, any eavesdropping efforts leave detectable
traces in the schemes of quantum cryptography. This sets the motivation
for designing protocols for QIA. 

As mentioned in the previous section, first ever protocol for QIA
was proposed by $\text{{\rm Cr�peau}}$ et al. \cite{CS_1995} in
1995, it was based on a scheme of oblivious transfer\textcolor{blue}{{}
}(OT) which is a cryptographic primitive. Interestingly, an extremely
strong result by Lo and Chau\textcolor{black}{{} \cite{LC_1997} }established
the impossibility of an unconditionally secure two-party OT. The pioneering
effort of $\text{{\rm Cr�peau}}$ et al. \cite{CS_1995} was followed
by a set of early works \cite{ZW_1998,DHHM_1999,B99} on QIA which
are briefly mentioned in the previous section, but not much activity
related to quantum or hybrid identity authentication happened in the
previous century. This situation changed considerably in the present
century. In the remaining part of this section, we will try to provide
a chronological history of the development of different protocols.
As the literature is vast, efforts will not be made to include all
works\footnote{For a complete list of papers published in the domain of quantum authentication
between 2009-2019, interested readers may see \cite{MD21}. We came
across this brief literature review work after completion of this
review. Interestingly, classification of QIA schemes made in \cite{MD21}
is consistent with the classification made in this work.}. However, the representative works mentioned here are expected to
provide a clear picture of the evolution of the schemes of quantum
and hybrid identity authentication. It's also expected to put light
on the fact that the different type of QKD schemes and other protocols
of cryptographic tasks, like continuous variable (CV) QKD (CV-QKD),
discrete variable (DV) QKD (DV-QKD), counterfactual QKD, semi-QKD,
etc., require different types of QIA schemes to ensure that the same
resources/devices can be used for performing the communication task
as well as the authentication task. Actually, applicability of earlier
protocols of QIA were primarily restricted to conventional conjugate
coding based on DV cryptographic schemes, but later the domain of
applicability has been considerably enhanced.
\begin{description}
\item [{2000:}] It was a very active year, and many (at least 10) new protocols
for QIA appeared. Most of them (all except 3 protocols of Ljunggren
et al. \cite{LBK_2000}) were Bell state based. Specifically,\textcolor{green}{{}
}Zeng et al. proposed a Bell-state based scheme for QIA \cite{ZG_2000}
in early 2000. A few months later, Zhang et al. proposed another Bell
state based protocol for QIA \cite{ZLG_2000}, where a pre-shared
Bell state is used as the pre-shared secret or equivalently as the
authentication key. In between the introduction of these two Bell
state based schemes, Jensen et al., \cite{JS_2000} proposed a bi-partite
entangled state based protocol for QIA\footnote{It was not specifically mentioned that the particles are maximally
entangled, but without affecting any result, we can consider this
as a Bell state based scheme for QIA.} and claimed that their protocol is secure even when Eve possesses
complete control over the classical and quantum channels. This work
was a generalization of 1999 work of Barnum \cite{B99} and it was
also interesting because it listed a set of open questions.\textcolor{green}{{}
}\\
Ljunggren et al. proposed a set of 5 protocols for authenticated QKD
\cite{LBK_2000} between two parties Alice and Bob. In their protocols
a third party, Trent was introduced as arbitrator. Three of the protocols
were single-photon \cite{MD21} based and the others were based on
Bell states. Interestingly, one of the entangled-state based protocols
was analogous to B92 protocol and the symmetry observed between schemes
of QKD and QIA reflected in that protocol allows us to think of transforming
other schemes of QKD into schemes for QIA or authenticated QKD. Further,
here the role of Trent is analogous to the role of controller in cryptographic
switch \cite{SOS+14,TP15}, and following the result of \cite{SOS+14,TP15},
one can easily generalize the three-party protocols of Ljunggren et
al.
\item [{2001:}] In an interesting work, Curty et al., proposed an optimal
scheme for quantum authentication of a classical message (CS01 protocol
hereafter ) \cite{CS_01}. It was optimal in the sense that it allowed
authenticated transmission of a two-bit classical message using a
one qubit authentication key (which can be viewed as an optimal pre-shared
secret). Interestingly, they classified the possible attacks on the
authentication schemes as: (i) no-message attack and (ii) message
attack. In CS01 scheme and many other schemes of message authentication,
perfect deterministic decoding of the classical message happens. This
implies that a protocol fails if and only if Bob accepts an unauthenticated
message as authenticated one. Curty et al. argued that the same may
happen in two different ways leading to two families of attacks. Specifically,
in no-message attack, Eve prepares a quantum state prior to the transmission
of a message from Alice and sends that to Bob aiming to pass the decoding
algorithm of Bob with certain probability $p_{f}$. This is referred
to as no-message attack, as it works in a situation where the legitimate
user Alice has not yet send a message. In contrast, in a message attack,
Eve intercepts the message sent by Alice and tries to obtain information
from that. Subsequently, she prepares a fake message using the information
obtained by interception and tries to pass the decoding algorithm
of Bob. This classification of attacks helped security analysis of
the subsequently developed schemes, but in the security analysis performed
in \cite{CS_01} and in a large number of subsequent papers, a noiseless
quantum channel has been considered which is an idealization and does
not correspond to the practical situations. 
\item [{2002:}] In CS01 protocol, a classical message was authenticated
with the optimal amount of quantum resources. However, that naturally
led to an interesting question like: How to authenticate a quantum
message with a quantum key? Can a qubit (which can be viewed as a
quantum message of minimum size) be authenticated using another qubit
(which can be viewed as pre-shared quantum secret of minimum size)?
These questions were addressed in Ref. \cite{CSPF_02}, where Curty
et al. established that a no-go theorem which can be stated as: A
qubit cannot be authenticated with just another qubit.\textcolor{green}{{}
}\\
T. Mihara proposed schemes for QIA and message authentication \cite{T.Mihara_2002}
using pre-shared entanglement. The QIA scheme proposed by them had
a trusted third party, while the message authentication scheme was
hybrid in nature as they used quantum resources along with Hash function.
Here, it may be noted that the approach adopted by them in designing
a message authentication scheme can be generalized easily and every
known scheme of QSDC can be used to design similar protocols. Interestingly,
it was soon established that this scheme uses quantum correlations
but doesn't exploit them and in principle the same task can be performed
by classically correlated states \cite{V03}. Further, quantum part
in the identity authentication scheme of Mihara \cite{T.Mihara_2002}
is vulnerable. To be precise, during entanglement distribution process
(cf. Step 2-3, of Section entitled ``Identifying Alice to Bob''
in Ref. \cite{T.Mihara_2002}) neither BB84 subroutine nor GV subroutine
\cite{sharma2016verification} is used and all the subsequent measurements
are performed using computational basis, consequently it's vulnerable
under man-in-the-middle attack. Of course, one can circumvent this
possibility using decoy qubits implementing one of the above mentioned
subroutines, as is routinely done in the schemes of QSDC (for example,
one may see the description of ping-pong protocol in Ref. \cite{P13}).
\item [{2004:}] Li et al. proposed a scheme of QIA that uses Bell states
\cite{LB_2004}. Specifically, a Bell state is pre-shared by Alice
and Bob as identification token. Subsequently, during the authentication
process, Alice creates another auxiliary Bell state to interact with
the ``identification token'' and sends the auxiliary information
to Bob, who performs some operations before performing Bell measurement
to complete the authentication procedure. As before, this Bell state
based scheme too does not exploit the nonlocal properties of Bell
states. Specifically, Bell inequality violation as a quantum resource
is not used for designing schemes for QIA.
\item [{2005:}] Zhou et al. proposed a scheme of QIA using quantum teleportation
and quantum entanglement swapping \cite{ZZZZ_2005}. This scheme claimed
to resolve the limitations of simple point-to-point QIA schemes as
entanglement swapping can be used to extend the distance over which
authentication can be done. This is a known trick and frequently used
in QKD to circumvent the limitation on allowed distance that arise
from the presence of noise and unavailability of quantum repeater.
However, such an approach often leads to a new security concern as
the device deployed in the middle to initiate entanglement swapping
needs to be trusted.\\
A way to compute efficiency of a scheme of authentication is to compare
how much pre-shared information it consumes during the authentication
process. The less it consumes, the more efficient is the scheme. In
view of this measure, an efficient QIA scheme was proposed by Peev
et al. \cite{PNM+05}.
\item [{2006:}] \textcolor{black}{Lee et al. proposed two quantum direct
communication protocols with user authentication \cite{LLY_2006}.
Here Alice can directly communicate a secret message to Bob without
any pre-shared secret. The protocols are constructed in two parts:
one part is dedicated for authentication and the other part is used
for direct communication. They introduced a third party, Trent as
more powerful than the other users and enabling him to authenticate
the other users participating in the communication scheme by distributing
tri-partite ``Greenberger-Horne-Zeilinger'' (GHZ) states. }\\
\textcolor{black}{Wang et al. proposed an authentication protocol
that allows a trusted third party to simultaneously authenticate multiple
users with the help of entanglement swapping and GHZ states \cite{WZT_2006}.
}\textcolor{green}{}\\
Zhang et al. \cite{ZZZX_2006} proposed a one-way QIA scheme by using
the mechanism of the ping-pong protocol \cite{ping-pong_BF_02} of
QSDC. This was a scheme of one-way QIA in the sense that in their
work Alice was considered as a reliable \emph{certification authority}
(CA) and Bob was considered as the user whose identity needs to be
verified when he would try to communicate with Alice. Technically,
many of the schemes proposed so far are one-way in this sense. However,
usually that's not explicitly mentioned as the capability of one-way
authentication implies that two-way authentication can be done by
using the same process two times. They showed that their scheme is
secure against individual attacks like impersonated fraudulent attack
(where Eve is unaware of the authentication key, but impersonates
Bob by forging a new qubit) and substitution fraudulent attack (where
to impersonate Bob, Eve tries to obtain the authentication key using
a new quantum channel and exploiting this channel and the qubits traveling
from Bob to Alice)\textcolor{blue}{.} Here it may be noted that message
attack and no-message attack are not always specified with these names,
but most of the attacks discussed so far can be viewed as message
attack or no-message attack.
\item [{2007:}] Zhang et al. \cite{ZLWS_2007} reported that Lee et al.'s
protocol published in 2006 is vulnerable under two types of attack
which can be performed by Trent. Zhang et al. also modified the protocol
of Lee et al. and proposed an improved protocol, which is free from
the limitations of the Lee et al.'s protocol. This type of cryptanalysis
showing a vulnerability of a scheme and subsequent development of
an improved scheme free from that vulnerability is very common in
the field of cryptography, and this may be treated as a representative
example, in the context of QIA.
\item [{2009:}] Using GHZ states Guang et al. proposed a scheme for simultaneous
multiparty QIA \cite{GY_2009}. The task performed by the scheme was
similar to the task performed by the 2006 protocol of Wang et al.
\cite{WZT_2006}. However, Guang et al.'s scheme was shown to be more
efficient compared to the Wang et al.'s scheme in the sense that in
comparison to Wang et al.'s scheme lesser quantum resources and lesser
quantum operations are needed in Guang et al.'s scheme. \textcolor{red}{}\\
Rass et al. \cite{RSG09} introduced an interesting idea, where the
authentication was performed at the end of the QKD scheme instead
of the usual practice of performing it in the beginning. To design
the scheme, they combined the ideas of QKD and quantum coin-flipping.
They also found tight bounds on the amount of pre-shared secrets required
for QIA.
\item [{2010:}] Dan et al. \cite{DXXN_2010} introduced a simple protocol
of QSDC with authentication. The protocol was based on Bell states
and the qubits were described as the polarization encoded qubits.
Though such specification is not always mentioned in the schemes of
authentication, wherever transmission of qubits is involved, it's
intrinsically assumed that the qubits used are photonic in nature
as the\textcolor{black}{{} }transmon\textcolor{black}{{} qubits} used
in superconductivity based quantum computing and other realization
of qubits are not easily transferable from one place to the other.
Dan et al. also established security of their scheme against a set
of typical attacks, e.g.,\emph{ }intercept and resend attack, Trojan
horse attack and entanglement attack.
\item [{2011:}] Huang et al. \cite{HZLZ_2011} introduced a Gaussian-modulated
squeezed state based CV-QIA protocol. This protocol and similar protocol
allow CV-QKD schemes to use CV states for authentication, too. This
is important as the use of DV quantum authentication in CV quantum
protocols decreases efficiency and increases complexity in the sense
that in such a case, devices/resources are required for performing
operations and measurements on both CV and DV states. 
\item [{2012:}] Gong et al. tried to utilize quantum one-way function to
design a scheme of QIA involving a trusted third party which was referred
to as a trusted server in their work \cite{GTZ12}. \\
A US patent (Patent number US 8,340,298 B2) was also given to MagiQ
Technologies for a hybrid scheme of QIA in a quantum cryptographic
network \cite{GB12}.\\
Skoric proposed a very interesting idea that combined quantum challenges
with classical physical unclonable functions (PUFs). Specifically,
the idea was that a classical PUF was challenged using a quantum state.
Because of nocloning theorem and collapse on measurement principle,
any eavesdropping effort would lead to detectable trace and a verifier
who sends the challenge via a quantum state would be sure about the
identity of the prover if he receives the expected quantum state in
response of the challenge. Here it may be noted that the idea of PUFs
were introduced in the beginning of this century by Pappu et al.,
\cite{PRT+02} as an entity embodied in a specific physical structure
which can be easily evaluated but cannot be characterized easily.
It may also be viewed as an entity that cannot be feasibly duplicated
because it's manufactured inherently with a large number of uncontrollable
degrees of freedom \cite{GHMSP_2014}. Interestingly, FPGAs, RFIDs
and many other devices can be used to implement PUFs and classical
authentication schemes based on PUFs \cite{ZPD+14}. However, our
interest is restricted to PUF based schemes for QIA. In what follows
we will observe that interest on such schemes are progressively increasing.
\item [{2013:}] Yang et al. proposed two Bell state based schemes of authenticated
direct secure quantum communication \cite{YTXZ_2013}. The protocols
involved a third party, Trent and they were hybrid in nature as Hash
functions were also used. The protocol allowed Alice and Bob to authenticate
each-other with the help of Trent using Bell states in accordance
with the user's secret identity sequence and one-way Hash functions
$h_{A}$ and $h_{B}$. 
\item [{2014:}] In controlled quantum teleportation, a controller Charlie
controls when Alice can teleport a quantum state to Bob \cite{P13,TVP15}.
To address the identification confirmation problem associated with
controlled teleportation, Tan et al. proposed two identity authentication
schemes \cite{TJ_2014} based on entanglement swapping. The scheme
is not efficient as it requires two (three) entanglement swapping
operations to confirm the identity of Charlie (Bob) by Alice. \\
Goorden et al. experimentally realized a scheme for quantum authentication
of a physical unclonable key \cite{GHMSP_2014}. The key used here
was essentially a PUF and classical in nature. It was authenticated
by illuminating the PUF with a pulse of light having less number of
photons compared to spatial degrees of freedom and a subsequent analysis
of the spatial shape of the reflected light. \\
Shi et al. proposed a quantum deniable authentication protocol based
on the property of unitary transformation and quantum one-way function
using GHZ state \cite{SZY_2014}. Their protocol can provide the\emph{
completeness} of authentication and \emph{deniability}\textcolor{blue}{\emph{.
}}\textcolor{black}{Here, completeness of}\textcolor{black}{\emph{
}}\textcolor{black}{an authentication protocol}\textcolor{black}{\emph{
}}\textcolor{black}{implies that the aimed receiver can always confirm
the authentication of the source of the message if both receiver and
sender follow the protocol. In turn, a slight deviation from the conditions
of traditional authentication schemes, a deniable authentication scheme
ensures that apart from the intended receiver, no one else can identify
sources of a given message, and the receiver cannot prove the source
of the message to a third party.}\textcolor{blue}{{} }\\
Yuan et al. \cite{YLP+14} proposed a feasible QIA scheme using a
single particle based ping-pong protocol, which is analogous to LM05
\cite{LM_05} protocol of QSDC. This protocol requires a relatively
lesser amount of quantum resources and simple-devices. \\
Fountain codes or rateless erasure codes are the codes having an interesting
property that using a set of $n$ source symbols one can generate
a sequence of encoding symbols having $m>n$ symbols and $m$ can
be as big as needed. Using this property in distributed scenario,
i.e., using distributed fountain code, Lai et al. proposed a scheme
for QIA \cite{LXO+14} in the context of quantum secret sharing. 
\item [{2015:}] Shi et al. \cite{SZZY_2015} improved their previous work
\cite{SZY_2014} by using a single photon source instead of a GHZ
state. The new protocol was more efficient and clearly required lesser
amount of quantum resources. It's easy to visualize as creation and
maintenance of single photon states are much easier than the same
for photonic GHZ states.
\item [{2016:}] Ma et al. proposed a teleportation-based CV-QIA protocol
\cite{MHBZ_2016} using two-mode squeezed vacuum state and coherent
state. Interestingly, CV-QKD protocols perform better compared to
their discrete variable counterparts in relatively short distances
\cite{XCQ+15}. Consequently, for a metropolitan network, CV-QKD equipped
with CV-QIA is expected to provide an efficient solution. \\
Rass et al. proposed a QIA scheme in connection to BB84 protocol \cite{RSS+16}.
The novelty of the scheme was derived from the introduction of a second
public channel which was considered to be disjoint from the channel
used for implementing the main BB84 scheme.\\
Security of the PUF based authentication schemes \cite{GHMSP_2014,S12}
was questioned in Ref. \cite{YGL+16} and it was shown that the earlier
works (see \cite{SMP13} and references therein) on the security analysis
of the PUF based QIA were incomplete in the sense that the earlier
works established security against challenge-estimation attacks only,
but challenge-estimation attacks can be outperformed by a cloning
attack. \\
The composable security of a QIA scheme was established by Hayden
et al. \cite{HLM16}. This was an important step for performing secure
quantum communication tasks over a network. Here it will be apt to
note that the composable security implies a real protocol implemented
should remain $\epsilon$-indistinguishable from an ideal protocol
for the same task. This allows us to obtain the information theoretic
security of the composite scheme as $\epsilon\leq\epsilon_{p}+\epsilon_{a}$-secure
if an $\epsilon_{p}$-secure cryptographic protocol is used with an
$\epsilon_{a}$-secure application \cite{R08}.
\item [{2017:}] Hong et al. presented a QIA protocol using single photon
states \cite{HCJ+17}. Being single photon based scheme, the protocol
was relatively less resource demanding and also more efficient as
it allowed authentication of two bits of authentication key sequence
using a single qubit. \\
Abulkasim et al. proposed an authenticated quantum secret sharing
(QSS) protocol \cite{AHKB_2017} based on Bell states. To obtain a
higher level of security, here participants encrypt the pre-shared
key before use. This trick is commonly used in quantum communication.
\\
Nikolopoulos et al. introduced a CV authentication scheme \cite{ND_2017}
using coherent state of light and wavefront-shaping and homodyne measurement
techniques. \\
 Portmann proposed a quantum authentication scheme using recycling
of the key \cite{Portmann_2017} and showed the security against insecure
noisy channels and shared secret key. They have also proved the number
of recycled key bits is optimal for some authentication protocols.
\item [{2018:}] Kang et al. proposed a mutual authentication protocol \cite{KHHYHM_2018}
that allows Alice and Bob to authenticate each other even in the presence
of an untrusted third party Trent. They used entanglement correlation
checking (using GHZ-like states) to ensure that the state distributed
by Trent is really entangled and random numbers to stop Trent from
directly deducing the key. Similar strategies can be used into the
other QIA protocols where a trusted third party is considered. Such
a trivial effort can yield a set of new protocols for QIA.\\
A US patent (Patent number US 9,887,976 B2) was given for multi-factor
authentication using quantum communication that involves two stages
for enrollment and identification \cite{HPT+18} with a computer system
to implement a trusted authority.
\item [{2019:}] Semi-quantum protocols are a class of protocols that extends
quantum security to some classical users. First semi-quantum protocol
for QKD was proposed by Boyer et al. in 2007 \cite{boyer2007quantum}.
Subsequently, semi-quantum schemes have been proposed for various
tasks \cite{STP17,thapliyal2018orthogonal,mishra2021quantum,asagodu2021quantum}.
However, no semi-quantum authentication protocol was proposed until
2019, although that was a basic requirement of semi-quantum QKD and
other semi-quantum cryptographic schemes. The gap was addressed by
Wen et al. in 2019 when they introduced a semi-quantum authentication
protocol \cite{WZGZ_2019} which is not required the quantum capability
of all the users. Specifically, it allowed quantum Alice (Bob) to
identify classical Bob (Alice).\textcolor{green}{{} }\\
Another facet of modern quantum cryptography also found its place
in the authentication protocol in this year when Liu et al. introduced
a QIA protocol \cite{LGXHZX_2019} which can be used in the counterfactual
QKD schemes. \\
Zheng et al. proposed a controlled QSDC (CQSDC) protocol with authentication
\cite{ZL_2019} using five-qubit cluster state. Though the scheme
was shown to be secure in the presence of noise, the five-qubit cluster
state used in this protocol is extremely difficult to prepare and
maintain. It's straight forward to extend Bell state based schemes
or GHZ state based schemes to the schemes that use $n$-partite entangled
states such that $n>3.$ However, such an extension is not usually
advantageous. Here it may be noted that this was not the first incident
when a five qubit cluster state was used to design a scheme for QIA.
In fact, in 2014, a QIA scheme using five qubit cluster state and
one-time pad was proposed \cite{CXZY_2014} and the same was cryptanalyzed
in 2016 \cite{GH_2016}. In addition, an entanglement swapping based
scheme using six-qubit cluster state was proposed in 2012. Further,
in 2015 an even larger entangled state (precisely six qubit state)
was used in an effort to design a scheme of the authenticated QSDC
scheme \cite{YSL+15}. None of these schemes involving multi-partite
entangled states are practically useful at the moment.
\item [{2020:}] Most of the protocols for QIA mentioned above have not
considered the effect of noise which is unavoidable in practical situations.
Addressing this issue directly, Qu et al. proposed a QIA protocol
based on three-photon error avoidance code \cite{QLW_2020}. This
scheme can effectively resist the interference of noise on information
transmission through quantum channel. \\
Zhang et al. presented a Bell state based protocol of QIA that relies
on entanglement swapping \cite{ZCSL_2020} and can provide security
against the attacking strategies of a semi-honest third party. Here
it may be noted that a semi-honest user strictly follows the protocol,
but tries to cheat remaining within that. Consequently, attacks of
a semi-honest third party are weaker compared to those of a dishonest
third party and it's always preferable to consider a dishonest third
party specially while dealing with schemes of quantum cryptography.
\item [{2021:}] Either existing protocols are cryptanalyzed or modified
to propose a newer version of QIA scheme in connection with a specific
application. Specifically, in \cite{XL21} and in \cite{DPM21}, QIA
schemes are discussed in the context of quantum private query and
QSDC, respectively. Further, an earlier proposed hybrid scheme of
authentication \cite{Z19} (which was an improved version of a single
photon based scheme proposed in 2017 \cite{HCJ+17}) was systematically
cryptanalyzed in Ref. \cite{GCM+21}. In Ref. \cite{GCM+21}, two
attacks are designed and it's shown that not only the protocols proposed
in Refs. \cite{HCJ+17,Z19}, a set of other existing protocols for
QIA are also vulnerable under these attacks. In addition, an intense
interest is observed in PUF based quantum authentication schemes \cite{JSR21,DKS+21}
and closely related applications \cite{PAA+21}.
\end{description}
Before we close this section, it would be apt to note that there are
a few variants of the authentication protocols. For example, we may
note that deniable authentication is a particular variant of authentication
protocol, which does not allow a receiver to prove the source of the
message to a third party. For example, a protocol for quantum deniable
authentication was proposed in \cite{SZY_2014}. Further, we must
note that most of the QIA protocols are discussed in the two-party
scenario involving sender Alice and receiver Bob, but there is a set
of schemes where a third party Trent (referred to as an authenticator)
is added. For example, using GHZ states \cite{LLY_2006}, a set of
two three-party schemes for QIA were proposed by Lee et al., but later
it was reported in Ref. \cite{ZLWS_2007} that Trent can eavesdrop
in both the protocols. Interestingly, in \cite{ZLWS_2007}, a solution
to circumvent the proposed attacks was also proposed. Further, in
\cite{LBK_2000}, a three party scenario where authentication was
done by an arbitrator Trent.

\section{Classification of the protocols for quantum identity authentication\label{sec:Classification-of-the} }

Classification of anything depends on the criterion selected. Here,
we will classify the existing schemes for QIA based on two criteria:
(i) quantum resources used to design the schemes and (ii) computational
or communication task inherently used to design the schemes. The classification
performed here is neither unique nor complete. One can, of course
select a different criterion to classify the schemes. The reason behind
classifying the schemes based on the above two criteria is to explore
the inherent symmetry among the existing protocols for QIA and thus
to create a framework for designing new protocols.

\subsection{Classification based on the quantum resources used\label{sec:resource-based-Classification}}

Based on quantum resources used, we can classify the existing schemes
of QIA simply into two classes: (a) schemes that use entangled states
\cite{LB_2004,WZT_2006,ZLG_2000,ZZZX_2006,ZCSL_2020,KHHYHM_2018,CXZY_2014,T.Mihara_2002}
and (b) schemes which don't use entangled states \cite{Z19,HCJ+17,SZZY_2015,YLP+14,DHHM_1999,ZWZ_2020}.

\subsubsection{Entangled state based schemes of QIA}

Bell states are the simplest entangled states, and they are easy to
prepare and maintain. Naturally, a large number of Bell state-based
schemes for QIA have been designed. Specifically, protocols are described
in Refs. \cite{ZLG_2000,ZG_2000,LBK_2000,CS_01,CSPF_02,T.Mihara_2002,LB_2004,ZZZZ_2005,ZZZX_2006,DXXN_2010,YTXZ_2013,AHKB_2017,ZCSL_2020}
using Bell states only. These protocols are different from each other
in many different aspects: some of them use Bell states to perform
entanglement swapping \cite{TJ_2014,WZT_2006}, some others use it
to design schemes for authentication in analogy to ping-pong protocol
for QSDC \cite{ping-pong_BF_02}. However, the beauty of Bell states
(more precisely, the advantages of nonlocality) is not yet explored
in its fullest. Specifically, one can also exploit the inherent symmetry
explored in this review to design device independent and semi-device
independent schemes for QIA. Further, simple QIA schemes based on
the violation of Bell inequality can also be devised. Of course implementation
of such schemes will be more demanding compared to the existing schemes,
but will have their own advantages. Further, there are existing schemes
of authentication, which uses multipartite entangled states. For example,
authentication schemes using three-qubit GHZ states are reported in
\cite{WZT_2006,LLY_2006,GY_2009,TJ_2014,SZY_2014} and three-qubit
GHZ-like states (which are also GHZ states) are used to design QIA
schemes in Refs. \cite{KHHYHM_2018,WZGZ_2019}. Also, in Ref. \cite{CXZY_2014},
five-qubit cluster states are used for identity authentication. Creation
and maintenance of such multi-partite entangled states are still challenging.
In addition, most of these entangled states based schemes of QIA require
quantum memory which is not available at the moment. 

\subsubsection{Schemes of QIA which don't use entanglement}

There are many schemes of QIA which uses separable states or more
precisely single photon states. For example, we may mention \cite{Z19,HCJ+17,SZZY_2015,YLP+14,DHHM_1999,ZWZ_2020}
and similar schemes.\textcolor{red}{{} }The advantage of these schemes
over the entangled state based schemes arises from the relative ease
of the preparation of single qubit states and the fact that most of
the entangled state based schemes need to store home qubits that requires
quantum memory which is not available, but the single qubit based
schemes generally does not require quantum memory. Consequently, these
protocols appear to be more feasible from the perspective of practical
implementation.\textcolor{blue}{{} }These protocols primarily use BB84
kind of encoding, where travel qubits are prepared in $Z=\{\vert0\rangle,\vert1\rangle\}$
basis or $X=\{|+\rangle,|-\rangle\}$ basis based on the pre-shared
authentication key and a predecided rule that connects bit value of
authentication key to the basis in which a travel qubit is to be prepared.
A particular scheme of this type is Zawadzki's \cite{Z19} scheme
for QIA. Recently, Guill�n et al. \cite{GCM+21} have shown that Zawadzki's
scheme is insecure in general and the claimed logarithmic security
of Zawadzki's scheme loses its advantage under a key space reduction
attack which is explicitly described by Guill�n et al. 

These kind of schemes often uses Hash function to ensure the security
of the authentication procedure. However, the use of Hash function
takes us outside the domain of quantum security, but at present confidence
of the cryptographers on a set of Hash functions is high. In what
follows, we propose a set of three new protocols for QIA in Section
\ref{sec:New-protocols}, out of which two protocols are single qubit
based, but not vulnerable under key space reduction attack and other
known strategies. 

\subsection{Classification of the protocols based on the computational or communication
tasks used to design the schemes \label{sec:task-based-Classification}}

In the section we will classify the existing protocols for QIA based
on the computational or communication tasks involved in the realization
of the scheme for QIA. To begin with, let us briefly describe the
schemes for QIA which are based on specific quantum communication
tasks.

\subsubsection{Protocols based on the schemes for QKD}

There exist many protocols of authentication, which are essentially
based on the existing schemes of QKD. In fact, all schemes for QKD
can in principle be reduced to schemes of QIA, provided there exists
a pre-shared secret key. Of course, one may need to slightly modify
the original protocol of QKD to reduce it to a scheme for QIA. For
example, such an effort was made in \cite{SKB11} where a scheme of
QIA was obtained by slightly modifying BB84 protocol of QKD. Lately,
a counterfactual QKD scheme was modified to propose a scheme for QIA
in \cite{LGXHZX_2019}. Further, in the early work of Du�ek\textcolor{black}{{}
et al. a classical-quantum hybrid scheme of authentication was used
where the quantum part was based on QKD. To be precise, they utilized
the fact that QKD schemes are essentially schemes for key amplification,
and thus if Alice and Bob start with a pre-shared sequence for identity
authentication and use that sequence only once to authenticate each
other and thus to generate a secure key using a scheme of QKD then
for a later run of identity authentication part they will be able
to use a part of the key generated by the scheme of QKD. Here, only
the refueling of the identity authentication sequence happens via
QKD, so this early effort does not really qualify as a scheme of QIA,
but it was definitely based on QKD. Similar use of QKD in the authentication
process is now common.}

\subsubsection{Protocols based on the schemes for QSDC and DSQC \label{subsec:Protocols-based-onQSDC}}

There are two types of schemes of secure quantum direct communication,
namely QSDC and DSQC. There exist many schemes of QSDC and DSQC (cf.
\cite{STP20,STP17,YSP14,BP12,SP14} and references therein) and each
of them provides a way to directly transfer a message from sender
Alice to receiver Bob in an unconditionally secure manner. Now, if
we assume that Alice and Bob have a pre-shared secret then Alice can
send a part or full of that secret to Bob by using a QSDC or DSQC
protocol and Bob can decipher the text and compare with his secret
to verify Alice's identity. Thus, in principle, every QSDC and DSQC
protocol can be converted to protocols for QIA. A specific example
of early QSDC protocol is ping-pong protocol, the same was used to
design a scheme for QIA by Zhang et al. in 2006 \cite{ZZZX_2006}
as an example of this class of QIA schemes. Here, we may note that
schemes for secure direct quantum communication that cannot be used
for long distance quantum communication due to noise can also be used
for QIA. Specifically, in the context of CV secure direct quantum
communication, this particular advantage of the schemes of secure
direct quantum communication was noted in Ref. \cite{PBM+08}. The
beauty of CV-QIA schemes (independent of whether it's designed using
a scheme of CV-QSDC or CV-DSQC) like the one hinted in \cite{PBM+08}
underlies in the fact that the existing infrastructure can be used
for the implementation of these schemes. 

LM05 \cite{LM_05} is a single photon based protocol for QSDC, which
is analogous to ping-pong protocol of QSDC. Using LM05 protocol, another
scheme of QIA was proposed by Yuan et al. in 2014 \cite{YLP+14} they
described it as a ping-pong scheme without entanglement, but it was
actually a modified LM05 scheme. Interestingly, neither Zhang et al.
nor Yuan et al. and the subsequent authors have recognized the essential
symmetry mentioned above which allows to transform all the existing
schemes for QSDC and DSQC to the schemes for QIA.

Reduction of schemes of QSDC or DSQC into the schemes for QIA often
involves some modifications which take the final scheme outside the
domain of QSDC or DSQC and in reality we obtain QSDC or DSQC inspired
schemes for QIA. For example, structure of Zhang et al.'s scheme \cite{ZLG_2000}
is analogous to the original ping-pong protocol, but it's different
in the sense that the travel qubit sent by Bob to Alice is not returned
by Alice further. Specifically, in Zhang et al. scheme, a Bell state
$\psi_{AB}^{+}$($=\frac{1}{\sqrt{2}}(\vert0\rangle_{A}\vert1\rangle_{B}+\vert1\rangle_{A}\vert0\rangle_{B})$)
is pre-shared by Alice and Bob, where the subscript $A(B)$ refers
to the qubit in possession of Alice (Bob). In the original ping-pong
protocol, Bob prepares the Bell state and sends a qubit (travel qubit)
to Alice, but this step may be considered as equivalent. Here Bob
prepares an arbitrary single qubit pure state $|\psi_{1}\rangle=\alpha|0\rangle+\beta|1\rangle$
and sends it to Alice, who performs a CNOT using her share of the
Bell sate as control qubit and $|\psi_{1}\rangle$ as the target qubit.
Subsequently, Alice returns the qubit indexed with subscript 1 (here
we get the analogy with ping-pong) and after receiving that Bob performs
a CNOT operation using his share of Bell state (qubit $B$) as control
qubit and qubit 1 as the target qubit. After that, Bob measures qubit
1 in a basis in which $|\psi_{1}\rangle$ is a basis element, if he
obtains $|\psi_{1}\rangle=\alpha|0\rangle+\beta|1\rangle$ then the
authentication succeeds. There are two specific reasons behind additional
attention being given to this type of QSDC based protocols. Firstly,
in what follows, we will propose new QSDC/DSQC based protocols and
secondly several QSDC/DSQC based protocols for QIA have recently been
proposed and those protocols may be viewed as protocols belonging
to the ping-pong family. For example, Li et al. proposed a QIA protocol
\cite{LB_2004} in a manner similar to Zhang et al.'s scheme with
a difference that in their protocol instead of single qubit state
$|\psi_{1}\rangle$ prepared and shared by Bob, Alice prepares an
auxiliary Bell state and performs a CNOT operation in which the first
qubit of auxiliary Bell state is used as the control qubit and her
qubit of the pre-shared Bell state (qubit $A$) as target. Then she
sends the auxiliary Bell pair to Bob who performs the CNOT operation
with the second qubit of the auxiliary pair as control qubit and qubit
$B$ as target qubit. If a subsequent Bell measurement performed by
Bob on the auxiliary Bell state finds it in the same state in which
it was prepared by Alice then authentication is considered to be successful.
The above two protocols require quantum memory, but quantum memory
devices are not yet available commercially. Even the lab-level quantum
memory devices available today require much development before those
can be used in real applications.\textcolor{blue}{{} }

Later, Zhang et al. introduced a one-way QIA protocol \cite{ZZZX_2006}
which is also similar to the protocol \cite{ZLG_2000}. This protocol
was followed by another protocol by Li et al. \cite{LC_2007} which
can be viewed as a modified version of their earlier protocol \cite{LB_2004}
as the auxiliary state is now prepared by Alice and Bob separately
in a single qubit state instead of Bell state.\textcolor{red}{{} }As
it appears from the brief description of the protocols, all the protocols
described in this section belongs to ping-pong family of protocols
for QIA. This family of protocols contains several other protocols,
including the single qubit based protocol of Yuan et al. \cite{YLP+14}.

\subsubsection{Protocols based on teleportation}

It's known that teleportation can be used for secure communication
provided we have an ideal (noise free) quantum channel \cite{LC99}.
This is a known issue and will be discussed in some more details in
Section \ref{subsec:Protocols-based-on-QECC}. Despite this issue
and associated issues involving entanglement concentration and/or
purification, several authors have proposed schemes for QIA based
on teleportation. For example, we may mention the schemes reported
in \cite{TJ_2014,MHBZ_2016,ZZZZ_2005} and the references therein.
Interestingly, even for CV-QIA a teleportation based scheme has been
proposed \cite{MHBZ_2016}. However, applicability of all these schemes
are limited by the requirement that in a noisy situation shared entanglement
will get corrupted and to obtain the desired entangled state from
that one would require to implement an entanglement purification or
concentration scheme which would require interaction between Alice
and Bob even before the authentication. Here it may be noted that
the above mentioned concern is also applicable to a large number of
entangled states based protocols for QIA (e.g., \cite{ZLG_2000} and
similar schemes) which do not explicitly use teleportation, but use
shared entanglement. This undesired interaction is a weakness of the
QIA schemes which are directly based on teleportation or shared entanglement.
However, clever tricks like one used in Barnum et al. \cite{BCC02}
(cf. Section \ref{subsec:Protocols-based-on-QECC} for a short discussion)
can help us to circumvent this limitation of the teleportation based
schemes for QIA. 

\subsubsection{Protocols based on quantum secret sharing}

Quantum secret sharing (QSS) schemes are also modified to design schemes
for QIA enabling authentication of the legitimate users individually
or simultaneously by a set of other users or trusted third party \cite{YWZ_08}.
In addition, a set of schemes for QSS with the feature of identity
authentication for better security have been proposed \cite{AHBR_16,AHKB_2017,shi2019useful,YC_21}.
A large variety of quantum resources and operations (e.g., Bell state,
phase-shift operation, GHZ state) have been used in these schemes.
In an interesting development, an effort was made in Ref. \cite{AHBR_16}
to simultaneously use the concept of QSS and QD \cite{BST+17,SKB+13}
to design protocols for mutual QIA. In multiparty version of the scheme
proposed by Abulkasim et al. Alice was considered to be the boss and
Bob and Charlie as her agents. In a later work \cite{GWWY_18}, a
vulnerability of the Abulkasim et al.'s scheme was reported and it
was shown that if the agents (Bob and Charlie) collude then the Abulkasim
et al.'s scheme becomes insecure. Responding to this criticism, Abulkasim
et al. proposed a modified version of their earlier scheme in \cite{AHE_18}
which was a QSS based scheme free from the collusion attacks of the
agents.

Protocols reviewed and classified in this section till now are all
derived from the protocols for different quantum communication tasks,
e.g., teleportation, QKD, QSDC and DSQC. In what follows, we will
show that a similar reduction is possible from the protocols for different
quantum computing tasks, too.

\section*{Protocols based on quantum computation tasks}

As mentioned above, existing protocols for different quantum computing
tasks can be modified to obtain schemes for QIA and the same has been
done since the early days of QIA schemes. In fact, the pioneering
effort for designing QIA by ${\rm Cr\acute{e}peau}$ et al. \cite{CS_1995}
involved OT. To begin with we will mention that, and will continue
to discuss the schemes designed (or can be designed) using the schemes
of other computing tasks, like blind quantum computing and quantum
private comparison.

\subsubsection{Protocols based on oblivious transfer}

As mentioned in Section \ref{sec:Introduction}, first protocol for
QIA was proposed by ${\rm Cr\acute{e}peau}$ et al. \cite{CS_1995}
in 1995 using OT. The scheme was aimed to check the identities of
the legitimate participants having a pre-shared common information
by comparing their mutual knowledge of this common information. This
pioneering effort to design a scheme of identity authentication whose
security is obtained from the attributes of quantum mechanics has
been followed by many others. In fact, security of all the protocols
for QIA mentioned in this review are obtained from the quantum features.
However, OT is hardly used as a resource to design schemes for QIA.

\subsubsection{Protocols based on blind quantum computing}

Blind quantum computing (BQC) was first proposed by Childs \cite{childs_2005},
his demand was that a user (client) with limited amount of quantum
resources may delegate a task to another user or quantum server having
full quantum capability or more quantum power, with the condition
that input and output of the client and the computational task performed
are kept in private, i.e., the server or the user with higher quantum
power remains blind about these information. Several schemes of BQC
with identity authentication have been proposed \cite{LLCZL_18,QLLSP_21,SCY_21},
but the capability for BQC in designing schemes of QIA is not yet
fully utilized. However, in Ref. \cite{LSG_17}, a scheme for blind
quantum signature- a task closely related to QIA has been proposed
using BQC. This may be mapped to a scheme of QIA and generalizing
the idea, efforts may be made to explore the possibility of designing
schemes for QIA using the existing and new schemes for BQC. This is
a strong possibility that needs to be explored properly. 

\subsubsection{Protocols based on quantum error detection or correction code \label{subsec:Protocols-based-on-QECC}}

A simple idea for performing authentication may be to start with a
shared entangled state(s) and a private quantum key which is to be
teleported by Alice to Bob using the shared entangled state for the
verification of her identity by Bob. Interestingly, the shared entanglement
can get corrupted due to noise and Alice and Bob may require to execute
a purity-testing protocol, which would check whether the shared entangled
state is in the desired state or not, but it would not try to correct
it as is done in the schemes of entanglement concentration and purification.
Barnum et al. \cite{BCC02}\textcolor{red}{{} }realized that such a
protocol for purity testing is expected to be interactive, but this
interactive nature is not desired in the schemes for QIA and quantum
message authentication specially when one-way authentication is desired.
Here comes the beauty of Barnum et al.'s \cite{BCC02} idea, they
first established that purity-testing schemes can be designed using
a family of quantum error-correcting codes (QECCs) having a specific
property that any Pauli error is detected by most of the codes in
the family. Subsequently, they showed that a scheme for \textcolor{black}{secure
non-interactive }quantum authentication  can be derived from any scheme
of purity-testing which can be obtained from a QECC. This is an interesting
work where QECCs were used for quantum authentication. Much later,
Qu et al. proposed a QIA protocol based on three-photon error avoidance
code \cite{QLW_2020,QLW19}. 

\subsubsection{Protocols based on quantum private comparison}

In 2017, Hong et al. proposed an efficient single qubit based protocol
for QIA \cite{HCJ+17} where decoy qubits were used for establishing
the security against eavesdropping. The protocol was efficient in
the sense that verification of two bits of authentication information
was performed using a single qubit. Later this protocol was cryptanalyzed
and improved by Zawadzki \cite{Z19}\textcolor{blue}{. }Zawadzki's
main criticism was that each run of Hong et al.'s protocol causes
some information leakage and an eavesdropper can continue collecting
such information in successive protocol runs. Zawadzki proposed an
improved version of Hong et al.'s protocol where Hash function was
also used. Interestingly, Guill�n et al. has recently shown that Zawadzki's
protocol is vulnerable against key space reduction attack \cite{GCM+21}.
Without going into a detailed discussion on that aspect, we would
like to note that in Zawadzki's protocol and many other protocols
for QIA the main task is actually to perform quantum private comparison
or to compute a function $f(A,B):f(A,B)=0$ if $A\neq B$ and $f(A,B)=1$
if $A=B$, using quantum resources without really disclosing the value
of $A$ and $B$ to a party who computes this function (of course,
some information will be logically obtained from the value of $f(A,B)).$
This is a standard problem of secure multiparty computation \cite{SaTP20,STP20,CGS02}
and the task is closely related to the socialist millionaire problem
\cite{SKB+13}. To visualize this let us look back on Zawadzki's protocol
in little more detail. In Zawadzki's protocol (and in many other protocols)
Alice and Bob share secret authentication key $k$. Alice generates
a random number $r_{A}$ and computes session secret $(s_{A}=H(k,r_{A}))$
using secret key ($k)$ and Hash function ($H$). Alice subsequently
sends the random number $r_{A}$ to Bob, who may receive a different
number $r_{B}$ and uses that to generate his own copy of session
secret $(s_{B}=H(k,r_{B}))$. The rest of the task is simply to use
quantum resources to compare whether $s_{A}=s_{B}$ or not or equivalently
to perform a quantum private comparison task by computing $f(s_{A},s_{B}).$
Equality of $s_{A}$ and $s_{B}$ would imply successful authentication.
Realization of this connection between the quantum private comparison
and QIA enables us to conclude that the existing schemes of quantum
private comparison and socialist millionaire problems can be reduced
to the protocols for QIA. For example, such protocols designed by
some of the present authors and reported in \cite{SKB+13,STP17,BST+17,SaTP20,thapliyal2018orthogonal}
can easily be reduced to the schemes for QIA. Interestingly, some
of such reduced schemes for QIA will even work in the semi-quantum
framework \cite{STP17,thapliyal2018orthogonal}, where all the users
need not have access to quantum resources. Here, it may be further
noted that capability of performing QD (simultaneous two-way communication
between Alice and Bob) or quantum conference \cite{BTS+18} ensures
the capability of performing direct secure quantum communication (QSDC
or DSQC) which can be used to perform QIA (as shown in Section \ref{subsec:Protocols-based-onQSDC}).
Thus, even without transforming the schemes of QD or quantum conference
into those of quantum private comparison, one can follow a relatively
direct route and design QSDC/DSQC based new schemes for QIA. 

\begin{figure}
\begin{centering}
\includegraphics[scale=0.5]{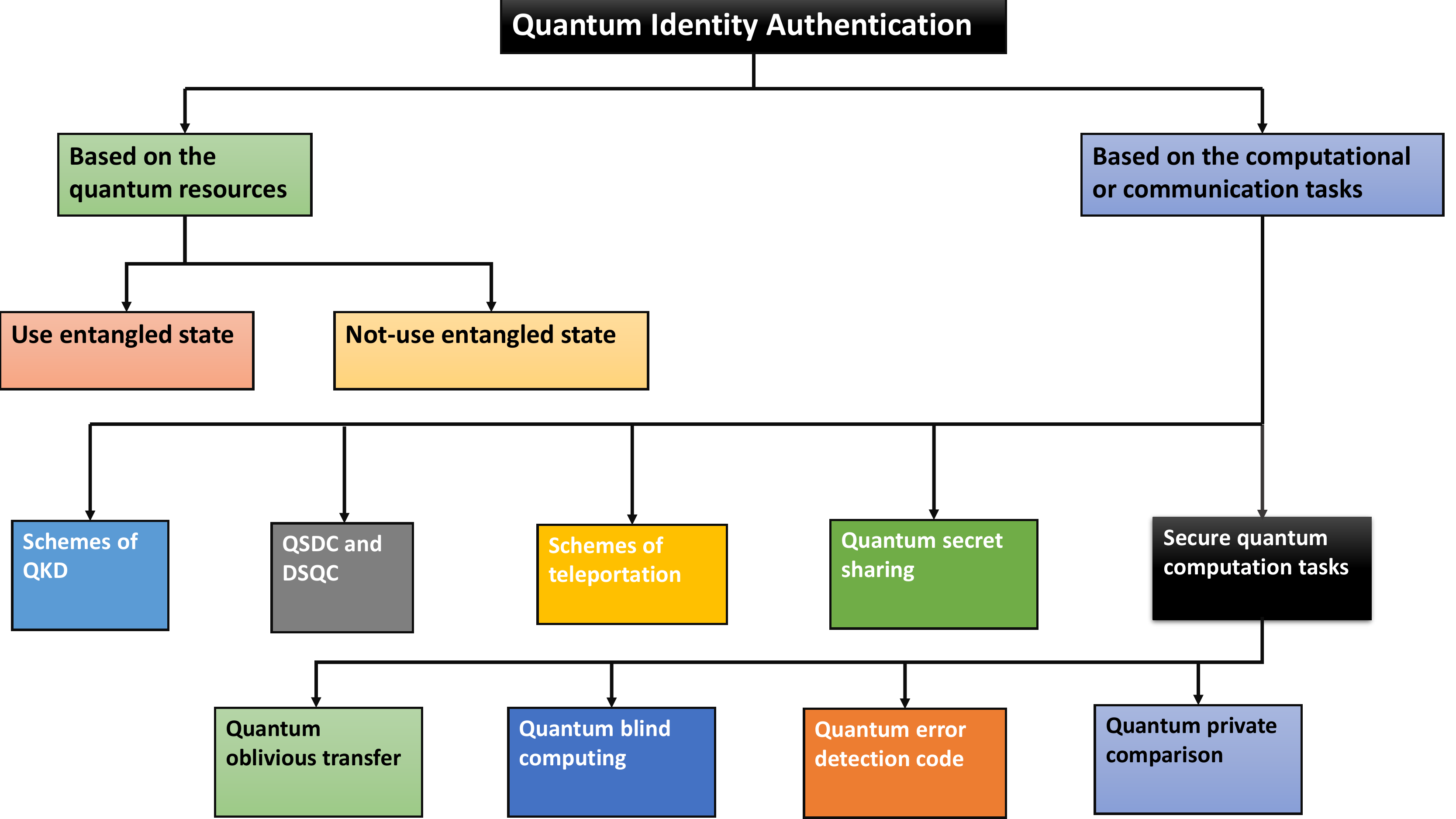}
\par\end{centering}
\caption{(Color online) Classification of the schemes for QIA. \label{fig:Classification}}
\end{figure}

Classification of the schemes for QIA performed in this section are
broadly summarized in Fig. (\ref{fig:Classification}). This figure
illustrates that QIA can be performed in various ways and using various
resources. Choice of a particular scheme would depend on the nature
of the task and the available resources. For instance, all the existing
schemes may be categorized on the basis of requirement of classical
channel, quantum memory and a third party as well as nature of pre-shared
key (classical or quantum) and the desired one-way or two-way authentication. 

\section{New protocols for quantum identity authentication\label{sec:New-protocols}}

In Section \ref{sec:A-short-chronological history} and \ref{sec:Classification-of-the},
we have hinted at various possibilities of designing new protocols
for QIA. As discussed above there are many possibilities of designing
new and modified protocols (of course some of them will be trivial
extension/modification of the existing protocols). To establish the
validity of our claim and to show that the symmetry among the protocols
appeared through the systematic review of the existing protocols can
really be exploited to design new protocols of QIA, here we propose
3 new protocols of QIA, all of which are based on the schemes for
secure direct quantum communication (i.e, QSDC and DSQC). These protocols
are to be viewed as representative protocols from the set of many
protocols which can be designed by using the protocols for other communication
and computation tasks. However, before we proceed, we must note that
the classification used here\textcolor{blue}{{} }is along the lines
of the literature of QIA, but in a strict sense the protocols described
below and in the existing literature do not qualify as direct communication
schemes. At best these schemes may be viewed as QSDC inspired or DSQC
inspired schemes for QIA. However, to avoid confusion among the readers
we are using here a relatively liberal definition of QSDC and DSQC
and classifying the schemes in line with the literature of QIA.

\subsection*{\textcolor{black}{New protocols based on QSDC }}

\textcolor{black}{In Section \ref{subsec:Protocols-based-onQSDC},
we have already mentioned that the schemes of secure direct quantum
communication can in principle be used to design schemes for QIA,
and a set of protocols for QIA has already been proposed using the
schemes for QSDC and DSQC. However, the identification of the existing
symmetry among these protocols allows us to construct new protocols.
As an example, here we propose two new protocols for QIA which are
based on QSDC. Let's begin with the first protocol.}

\subsection*{\textcolor{black}{Protocol 1:}}

\textcolor{black}{This QSDC based protocol for QIA is distinct but
analogous to the schemes proposed in \cite{CXZY_2014,YLP+14}. In
what follows, this QSDC inspired protocol is referred to as Protocol
1 and its steps are described as Step $1.j$ (in general Step $i.j:i,j\in\{1,2,\cdots\}$
would refer to $j^{th}$ step of the newly proposed $i^{th}$ protocol).
In this protocol, it's assumed that Alice and Bob posses a pre-shared
authentication key $K=\{k_{1},k_{2},k_{3},\cdots,k_{2n}\}$, and the
protocol is described in the following steps. }
\begin{itemize}
\item \emph{Encoding mode}
\end{itemize}
\begin{table}[H]
\caption{\label{tab:The-preparation-method-1}The preparation method of the
decoy sequence (Protocol 1)}

\centering{}%
\begin{tabular*}{16.5cm}{@{\extracolsep{\fill}}ccc}
\hline 
$(2i)^{th}$ bit value: $k_{2i}$ & 0 & 1\tabularnewline
$i^{th}$ qubit of decoy sequence: $K_{d_{i}}$ & $|0\rangle$~or~$|1\rangle$ & $|+\rangle$~or~$|-\rangle$\tabularnewline
\hline 
\end{tabular*}
\end{table}
\begin{description}
\item [{\textcolor{black}{Step~1.1:}}] \textcolor{black}{Alice prepares
an ordered $n$-qubit decoy sequence $K_{d}$ depending upon the bits
of the pre-shared key $K$. If $k_{2i}=0$, $i^{th}$ particle of
$K_{d}$ is prepared in the state $\vert0\rangle$ or $\vert1\rangle$,
otherwise the $i^{th}$ particle is prepared in the state $\vert+\rangle$
or $\vert-\rangle$ with only Alice knowing the exact states of the
particles in $K_{d}$.}
\item [{\textcolor{black}{Step~1.2:}}] \textcolor{black}{Alice prepares
another particle sequence $K_{a}$ to be used as authentication qubits.
The $i^{th}$ particle of $K_{a}$ is in state $\vert0\rangle$ if
$k_{2i-1}\oplus k_{2i}=0$, else in state $\vert-\rangle$. Alice
inserts the authentication qubits of $K_{a}$ into the sequence $K_{d}$
to form an enlarged sequence $K_{A}$ according to the rule that if
$k_{2i-1}\oplus k_{2i}=0$, then Alice puts the $i^{th}$ particle
of $K_{a}$ after the $i^{th}$ particle in $K_{d}$ or else, puts
the $i^{th}$ particle of $K_{a}$ before the $i^{th}$ particle in
$K_{d}$. Alice then sends the enlarged sequence $K_{A}$ to Bob.}
\end{description}
\begin{table}[h]
\caption{\label{tab:The-preparation-method}The preparation method of the authentication
sequence (Protocol 1)}

\centering{}%
\begin{tabular*}{16.5cm}{@{\extracolsep{\fill}}ccc}
\hline 
Additional modulo 2: $k_{2i-1}\oplus k_{2i}$ & 0 & 1\tabularnewline
$i^{th}$ qubit of authentication sequence: $K_{a_{i}}$ & $|0\rangle$ & $|-\rangle$\tabularnewline
\hline 
\end{tabular*}
\end{table}
\begin{itemize}
\item \emph{Decoding mode}
\end{itemize}
\begin{table}[h]
\caption{\label{tab:Measurement-basis-used}Measurement basis used by Bob to
get the result for the authentication sequence (Protocol 1) }

\centering{}%
\begin{tabular*}{16.5cm}{@{\extracolsep{\fill}}ccc}
\hline 
Additional modulo 2: $k_{2i-1}\oplus k_{2i}$ & 0 & 1\tabularnewline
Measurement basis: $B_{i}=\{B_{z},B_{x}\}$ & $B_{z}=\{|0\rangle,|1\rangle\}$ & $B_{z}=\{|+\rangle,|-\rangle\}$\tabularnewline
Measurement result: & $|0\rangle$ & $|-\rangle$\tabularnewline
\hline 
\end{tabular*}
\end{table}
\begin{description}
\item [{\textcolor{black}{Step~1.3:}}] \textcolor{black}{According to
the pre-shared key $K$, Bob calculates the position and the basis
used for the decoy qubits and authentication qubits by Alice then
Bob measures the decoy qubits $K_{d}$ and authentication qubits $K_{a}$
sequentially applying the rules that, when $k_{2i}=0$, Bob will use
the $Z$-basis ($\{\vert0\rangle,\vert1\rangle\}$) to measure the
particles of $K_{d}$, else he will use $X$-basis ($\{\vert+\rangle,\vert-\rangle\}$)
and measures the authentication qubit $K_{a}$ according to the value
of }\textbf{\textcolor{black}{$k_{2i-1}\oplus k_{2i}$}}\textcolor{black}{{}
and matches them with the expected outcomes which is described in
Table (\ref{tab:Measurement-basis-used}). Bob calculates the QBER
value using the measurement result of decoy sequence $K_{d}$, if
QBER is lesser than a tolerable limit, it confirms that the channel
is secure. If the measurement result of authentication sequence $K_{a}$
mismatches by greater than the tolerable limit, then Bob discards
the protocol, else Bob confirms the identity of Alice.}
\item [{\textcolor{black}{Step~1.4:}}] \textcolor{black}{Bob prepares
the decoy sequence $K_{d}^{\prime}$ using the same method described
in Step 1.1 and also prepares the authentication qubit sequence $K_{a}^{\prime}$
applying same encoding rules and uses this rule to form an enlarged
sequence $K_{B}$. Bob then sends the enlarged sequence $K_{B}$ to
Alice.}
\item [{\textcolor{black}{Step~1.5:}}] \textcolor{black}{Alice separates
the decoy sequence from the authentication qubits and }confirms the
identity of Bob in a manner similar to that mentioned in Step 1.3.
\end{description}

\subsection*{\textcolor{black}{Protocol 2:}}

\textcolor{black}{As in the previous protocol, Alice and Bob are assumed
here to pre-share an authentication key $K=\{k_{1},k_{2},k_{3},\cdots,k_{4n}\}$
with the only difference with the previous protocol is that the length
of the pre-shared key is now $4n$ which is double in comparison to
the previous scheme.}
\begin{itemize}
\item \emph{Encoding mode}
\end{itemize}
\begin{table}[h]
\caption{\label{tab:The-preparation-method-2}The preparation method of the
authentication sequence (Protocol 2)}

\centering{}%
\begin{tabular*}{16.5cm}{@{\extracolsep{\fill}}ccc}
\hline 
Additional modulo 2: $k_{2i-1}\oplus k_{2i}$ & 0 & 1\tabularnewline
Bit value: $k_{2i}$ & %
\begin{tabular}{cc}
0 & 1\tabularnewline
\end{tabular} & %
\begin{tabular}{cc}
0 & 1\tabularnewline
\end{tabular}\tabularnewline
Prepared authentication state: $K_{A_{i}}$ & %
\begin{tabular}{cc}
$|0\rangle$ & $|1\rangle$\tabularnewline
\end{tabular} & %
\begin{tabular}{cc}
$|+\rangle$ & $|-\rangle$\tabularnewline
\end{tabular}\tabularnewline
\hline 
\end{tabular*}
\end{table}
\begin{description}
\item [{Step~2.1:}] \textcolor{black}{Alice prepares an ordered $n$-particle
sequence $K_{A}$ using the first half of the pre-shared authentication
key (2$n$ bits) with the condition that when $k_{2i-1}\oplus k_{2i}=0$,
then Alice prepares the particle state as $\vert0\rangle$ and $\vert1\rangle$
respectively for $k_{2i}=0$ and $k_{2i}=1$. Further, when $k_{2i-1}\oplus k_{2i}=1$,
then Alice prepares the particle state as $\vert+\rangle$ and $\vert-\rangle$
respectively for $k_{2i}=0$ and $k_{2i}=1$. Alice then sends the
ordered particle sequence $K_{B}$ to Bob. }
\item [{Step~2.2:}] \textcolor{black}{Bob performs a measurement on the
$i^{th}$ particle of the ordered sequence $K_{B}$ in the $Z$-basis
($\{\vert0\rangle,\vert1\rangle\}$) if $k_{2i-1}\oplus k_{2i}=0$
or else use $X$-basis ($\{\vert+\rangle,\vert-\rangle\}$) . Bob
forms a classical bit sequence $S_{B}$ by using the knowledge that
quantum states $\vert0\rangle$ and $\vert+\rangle$ represent 0 while
$\vert1\rangle$ and $\vert-\rangle$ represent 1. Bob compares the
$i^{th}$ bit of $S_{B}$ with $k_{2i}$ and for an ideal case it
should be $K_{B}=k_{2i}$. If the error rate is less than the tolerable
limit, then Bob confirms the identity of Alice.}
\end{description}
\begin{itemize}
\item \emph{Decoding mode}
\end{itemize}
\begin{table}[h]
\caption{\label{tab:Measurement-basis-used-2}Measurement basis used by Bob
to get the result for the authentication sequence (protocol 2)}

\centering{}%
\begin{tabular*}{16.5cm}{@{\extracolsep{\fill}}ccc}
\hline 
Additional modulo 2: $k_{2i-1}\oplus k_{2i}$ & 0 & 1\tabularnewline
Bit value: $k_{2i}$ & %
\begin{tabular}{cc}
0 & 1\tabularnewline
\end{tabular} & %
\begin{tabular}{cc}
0 & 1\tabularnewline
\end{tabular}\tabularnewline
Measurement basis: $B_{i}=\{B_{z},B_{x}\}$ & $B_{z}=\{|0\rangle,|1\rangle\}$ & $B_{z}=\{|+\rangle,|-\rangle\}$\tabularnewline
Measurement result: & %
\begin{tabular}{cc}
$|0\rangle$ & $|1\rangle$\tabularnewline
\end{tabular} & %
\begin{tabular}{cc}
$|+\rangle$ & $|-\rangle$\tabularnewline
\end{tabular}\tabularnewline
\hline 
\end{tabular*}
\end{table}
\begin{description}
\item [{Step~2.3:}] \textcolor{black}{Bob performs the Step 2.1 similar
to that of Alice with the rest half of the pre-shared authentication
key. }
\item [{Step~2.4:}] Alice performs as the Step 2.2 to confirm the identity
of Bob.
\end{description}
As mentioned above Protocols 1 and 2 are just examples of the protocols
which can be designed using the schemes for QSDC. Following the same
line, other existing schemes of QSDC can also be modified to obtain
the schemes for QIA. The security of these schemes of QIA will be
analyzed in Section \ref{sec:Security-analysis}. Before we proceed
to security analysis, let us propose another scheme for QIA. 

\subsection*{\textcolor{black}{New protocol based on controlled DSQC}}

We know that QSDC is one facet of the secure direct quantum communication,
and DSQC is the other facet of it. We have already presented two schemes
based on QSDC, let us know provide an example establishing that schemes
of DSQC (more precisely a controlled version of DSQC, i.e., Controlled
DSQC (CDSQC)) can also be modified to design schemes for QIA. Before
we explicitly describe our $3^{rd}$ protocol, we would like to briefly
describe the basic idea behind the designing of this protocol.

\subsubsection*{Basic idea\label{subsec:Basic-idea}}

This protocol is based on the properties of Bell state and entanglement
swapping. The corresponding relations between the Bell state and the
pre-shared authentication key is as follows:

\begin{equation}
\begin{array}{c}
00:\vert\phi^{+}\rangle=\frac{1}{\sqrt{2}}(\vert00\rangle+\vert11\rangle),\\
01:\vert\phi^{-}\rangle=\frac{1}{\sqrt{2}}(\vert00\rangle-\vert11\rangle),\\
10:\vert\psi^{+}\rangle=\frac{1}{\sqrt{2}}(\vert01\rangle+\vert10\rangle),\\
11:\vert\psi^{-}\rangle=\frac{1}{\sqrt{2}}(\vert01\rangle-\vert10\rangle).
\end{array}\label{eq:1}
\end{equation}
Also, the relations between the corresponding Pauli operations and
pre-shared authentication key\textcolor{red}{{} }is as follows:

\begin{equation}
\begin{array}{c}
00:I=\vert0\rangle\langle0\vert+\vert1\rangle\langle1\vert,\\
01:\sigma_{x}=\vert0\rangle\langle1\vert+\vert1\rangle\langle0\vert,\\
10:i\sigma_{y}=\vert0\rangle\langle1\vert-\vert1\rangle\langle0\vert,\\
11:\sigma_{z}=\vert0\rangle\langle0\vert-\vert1\rangle\langle1\vert.
\end{array}\label{eq:2}
\end{equation}

Now, let us briefly describe the idea behind this protocol. Suppose
Alice and Bob have previously shared authentication key (bit string)
of a certain length. Consider the bit string as 10, so Alice and Bob
prepare the Bell states as per (\ref{eq:1}) 

\begin{equation}
\begin{array}{lcl}
\vert\psi^{+}\rangle & = & \frac{1}{\sqrt{2}}(\vert01\rangle_{12(34)}+\vert10\rangle_{12(34)}),\end{array}\label{eq:3}
\end{equation}
With the subscripts 1,2 and 3,4 representing the particles of Alice
and Bob, respectively. Alice (Bob) sends Particle 2 (4) to (semi-honest)
Charlie, who acts as the third party authenticator. Charlie prepares
one of the following quantum states

\begin{equation}
\begin{array}{c}
\vert\pm\rangle=\frac{1}{\sqrt{2}}(\vert0\rangle_{5}\pm\vert1\rangle_{5}).\end{array}\label{eq:4}
\end{equation}
Let's assume, Charlie chooses $\vert-\rangle$ state. The combined
state of Alice, Bob and Charlie can be expressed as 
\[
\begin{array}{lcl}
\vert\psi^{+}\rangle_{12}\otimes\vert\psi^{+}\rangle_{34}\otimes\vert-\rangle_{5} & = & \frac{1}{2\sqrt{2}}[(\vert01\rangle+\vert10\rangle)_{12}\otimes(\vert01\rangle+\vert10\rangle)_{34}\otimes(\vert0\rangle-\vert1\rangle)_{5}].\end{array}
\]

Charlie performs a CNOT operation with Particle 5 acting as control
qubit and the target qubit chosen randomly between Particle 2 or 4.
Charlie then sends the Particle 2 to Bob and Particle 4 to Alice.
Meanwhile, Alice and Bob perform $i\sigma_{y}$ operator (according
to their secret key $S_{AB}=\{10\}$ as shown in (\ref{eq:2}) on
their Particles 1 and 3 respectively. Let us suppose that for the
CNOT gate, the target is particle 2 then the final joint state after
performing the CNOT operation and Pauli operation can be written as,

\[
\begin{array}{lcl}
|\Psi\rangle_{12345} & = & \frac{1}{2\sqrt{2}}[\vert11110\rangle-\vert10111\rangle-\vert11000\rangle+\vert10001\rangle\\
 & - & \vert00110\rangle+\vert01111\rangle+\vert00000\rangle-\vert01001\rangle]_{12345},
\end{array}
\]
This equation can be rearranged and written as

\begin{equation}
\begin{array}{lcl}
|\Psi\rangle_{14235} & = & \frac{1}{2\sqrt{2}}[(\vert\phi^{+}\rangle_{14}\otimes|\phi^{+}\rangle_{23}+\vert\phi^{-}\rangle_{14}\otimes|\phi^{-}\rangle_{23}\\
 & - & \vert\psi^{+}\rangle_{14}\otimes|\psi^{+}\rangle_{23}-\vert\psi^{-}\rangle_{14}\otimes|\psi^{-}\rangle_{23})\otimes|0\rangle_{5}\\
 & + & (-\vert\phi^{+}\rangle_{14}\otimes|\psi^{+}\rangle_{23}+\vert\phi^{-}\rangle_{14}\otimes|\psi^{-}\rangle_{23}\\
 & + & \vert\psi^{+}\rangle_{14}\otimes|\phi^{+}\rangle_{23}-\vert\psi^{-}\rangle_{14}\otimes|\phi^{-}\rangle_{23})\otimes|1\rangle_{5}.
\end{array}\label{eq:5}
\end{equation}

Alice and Bob then perform the Bell measurements on the particles
which they have right now (i.e., 1 and 4 with Alice and 2 and 3 with
Bob) and send to Charlie the classical bits corresponding to their
measurement outcomes (as per (\ref{eq:1})). Charlie then measures
the Particle 5 in the computational basis $(\{\vert0\rangle,|1\rangle\})$
to get either the state $|0\rangle$ or $|1\rangle$ with $\frac{1}{2}$
probability. Charlie then performs the XOR operation on the classical
bits sent by Alice and Bob. The list of possible outcomes as per the
measurements performed by Alice, Bob and Charlie for the above example
has been shown in Table (\ref{tab:7}).

\begin{table}[h]
\caption{\label{tab:7}The possible measurement results by all the parties
in Protocol 3 for QIA.\textcolor{red}{{} }}

\centering{}%
\begin{tabular*}{16.5cm}{@{\extracolsep{\fill}}ccc}
\hline 
Alice and Bob's possible combination result & Charlie's result & Additional modulo 2 \tabularnewline
\hline 
$|\phi^{+}\rangle_{14}\oplus|\phi^{+}\rangle_{23}$ & $|0\rangle_{5}$ & $00\oplus00=00$\tabularnewline
$\vert\phi^{-}\rangle_{14}\oplus|\phi^{-}\rangle_{23}$ & $|0\rangle_{5}$ & $01\oplus01=00$\tabularnewline
$\vert\psi^{+}\rangle_{14}\oplus|\psi^{+}\rangle_{23}$ & $|0\rangle_{5}$ & $10\oplus10=00$\tabularnewline
$\vert\psi^{-}\rangle_{14}\oplus|\psi^{-}\rangle_{23}$ & $|0\rangle_{5}$ & $11\oplus11=00$\tabularnewline
$\vert\phi^{+}\rangle_{14}\oplus|\psi^{+}\rangle_{23}$ & $|1\rangle_{5}$ & $00\oplus10=10$\tabularnewline
$\vert\phi^{-}\rangle_{14}\oplus|\psi^{-}\rangle_{23}$ & $|1\rangle_{5}$ & $01\oplus11=10$\tabularnewline
$\vert\psi^{+}\rangle_{14}\oplus|\phi^{+}\rangle_{23}$ & $|1\rangle_{5}$ & $10\oplus00=10$\tabularnewline
$\vert\psi^{-}\rangle_{14}\oplus|\phi^{-}\rangle_{23}$ & $|1\rangle_{5}$ & $11\oplus01=10$\tabularnewline
\hline 
\end{tabular*}
\end{table}
We can see from the Table \ref{tab:7} that if the Charlie gets the
state $|0\rangle$ then XOR of the classical bits sent by Alice and
Bob will always be 00. For the case in which Charlie gets the state
$|1\rangle$ then XOR of the classical bits sent by Alice and Bob
will always be 10. Any deviation from from the mentioned outcomes
reveals the presence of an Eve. 

\subsection*{Protocol 3:}

Here, two parties Alice and Bob wish to verify themselves as authenticated
users through a third party Charlie. In this protocol, Alice and Bob
have a pre-shared sequence of a classical secret key $K_{AB}=\{k_{1}^{1}k_{2}^{1},k_{1}^{2}k_{2}^{2},k_{1}^{3}k_{2}^{3},\cdots,k_{1}^{i}k_{2}^{i},\cdots,k_{1}^{n}k_{2}^{n},\}$.
The steps involved in the authentication are as follows: 
\begin{description}
\item [{Step~3.1:}] Alice and Bob separately prepare a sequence of $n$
Bell states ${A}_{12}$ and ${B}_{34}$ respectively as per the pre-shared
key $K_{AB}$ and rule mentioned in (\ref{eq:1}). 
\begin{equation}
\begin{array}{c}
A=\{\vert A\rangle_{12}^{1},|A\rangle_{12}^{2},|A\rangle_{12}^{3},\cdots,|A\rangle_{12}^{i},\cdots,|A\rangle_{12}^{n}\}\\
B=\{\vert B\rangle_{34}^{1},|B\rangle_{34}^{2},|B\rangle_{34}^{3},\cdots,|B\rangle_{34}^{i},\cdots,|B\rangle_{34}^{n}\}
\end{array}
\end{equation}
The subscripts 1, 2, 3 and 4 are just used to distinguish the particles
in the sequences. Ideally, $A=B$. 
\item [{Step~3.2:}] Alice and Bob divide each state of their Bell states
into two ordered sequences of $n$ particles, each with the first
particle of each Bell state forming one sequence while the second
particle of Bell states forming the other sequence.
\end{description}
\begin{equation}
\begin{array}{c}
S_{1}=\{s_{1}^{1},s_{1}^{2},s_{1}^{3},\cdots,s_{1}^{i},\cdots,s_{1}^{n}\}\\
S_{2}=\{s_{2}^{1},s_{2}^{2},s_{2}^{3},\cdots,s_{2}^{i},\cdots,s_{2}^{n}\}\\
S_{3}=\{s_{3}^{1},s_{3}^{2},s_{3}^{3},\cdots,s_{3}^{i},\cdots,s_{3}^{n}\}\\
S_{4}=\{s_{4}^{1},s_{4}^{2},s_{4}^{3},\cdots,s_{4}^{i},\cdots,s_{4}^{n}\}
\end{array}
\end{equation}

Here $S_{1}$ is the sequence of the first particles of all Bell states
in $A$ and $S_{2}$ is the sequence of the second particles of all
Bell states in $A$. Similarly, $S_{3}$ is the sequence of the first
particles of all Bell states in $B$ and $S_{4}$ is the of sequence
the second particles of all Bell states in $B$. Alice and Bob respectively
hold the particle sequence $S_{1}$ and $S_{3}$ with themselves.
Further, Alice (Bob) randomly inserts the decoy particles $d_{a}$
($d_{b}$) into the sequence $S_{2}$ $(S_{4})$ to form an enlarged
sequence $S_{2}^{\prime}$ $(S_{4}^{\prime})$. Alice (Bob) sends
the sequence $S_{2}^{\prime}$ $(S_{4}^{\prime})$ through the quantum
channel to Charlie.
\begin{description}
\item [{Step~3.3:}] Charlie receives the sequence $S_{2}^{\prime}$ and
$S_{4}^{\prime}$ and performs the security tests using the decoy
particles. After the successful security tests, Charlie removes the
decoy particles $d_{a}$ ($d_{b}$) to get the original sequence $S_{2}$
$(S_{4})$. Further, Charlie also prepares a sequence of n qubits
$S_{5}=\{s_{5}^{1},s_{5}^{2},s_{5}^{3},\cdots,s_{5}^{i},\cdots,s_{5}^{n}\}$
with each qubit arbitrary in the state $\vert+\rangle$ or $\vert-\rangle$.
Charlie then performs a CNOT operation with the particles of sequence
$S_{5}$ acting as control qubit and the particles of sequence $S_{2}$
or $S_{4}$ as the target qubit randomly.
\item [{Step~3.4:}] Charlie then prepares decoy particles $d_{c}$ ($d_{d}$)
and inserts them randomly into sequence $S_{2}$ and $S_{4}$ to form
an enlarged sequence $S_{2}^{\prime}$ $(S_{4}^{\prime})$. Charlie
sends the sequence $S_{2}^{\prime}$ $(S_{4}^{\prime})$ to Bob (Alice).
After successfully performing the security checks with the decoy particles
and removing them, Alice (Bob) now holds the sequences $S_{1}$ and
$S_{4}$ $(S_{2}$ and $S_{3}).$ 
\item [{Step~3.5:}] Alice (Bob) performs the Pauli operation on the particles
of the sequence $S_{1}$ ($S_{3}$) corresponding to pre-shared classical
key as per (\ref{eq:2}). After the Pauli operation the sequences
change into $S_{1}^{*}$ and $S_{3}^{*}$. Now, Alice performs the
Bell measurement on the particles of sequence $S_{1}^{*}$ and $S_{4}$
and notes the measurement outcome as classical sequence $R_{14}=\{r_{14}^{1},r_{14}^{2},\cdots,r_{14}^{i},\cdots,r_{14}^{n}\}$
as per (\ref{eq:1}). Bob also performs the Bell measurement on the
particles of sequence $S_{2}$ and $S_{3}^{*}$ and notes the measurement
outcome in form of classical sequence $R_{23}=\{r_{23}^{1},r_{23}^{2},\cdots,r_{23}^{i},\cdots,r_{23}^{n}\}$.
Alice and Bob send their classical sequences $R_{14}$ and $R_{23}$
to Charlie. Meanwhile, Charlie also measures the sequence $S_{5}$
in the computational basis \textbf{$\{\vert0\rangle,|1\rangle\}$}
to get a sequence $R_{5}=\{r_{5}^{1},r_{5}^{2},\cdots,r_{5}^{i},\cdots,r_{5}^{n}\}$,
with $r_{5}^{i}=0$ or $r_{5}^{i}=1$ corresponding to the situations
$|0\rangle$ and $|1\rangle$.
\item [{Step~3.6:}] Now, the third party Charlie has three classical sequences
namely $R_{14}=\{r_{14}^{1},r_{14}^{2},\cdots,r_{14}^{i},\\*\cdots,r_{14}^{n}\}$,
$R_{23}=\{r_{23}^{1},r_{23}^{2},\cdots,r_{23}^{i},\cdots,r_{23}^{n}\}$
and $R_{5}=\{r_{5}^{1},r_{5}^{2},\cdots,r_{5}^{i},\cdots,r_{5}^{n}\}$.
Charlie then performs the XOR on the bit strings $R_{14}$ and $R_{23},$
that are in the same positions. For the value of $r_{5}^{i}$ as $0$
($1$), if the XOR outcome is $r_{14}^{i}\oplus r_{23}^{i}=00$ ($r_{14}^{i}\oplus r_{23}^{i}=10$),
then the identity of Alice and Bob is authenticated by Charlie.
\item [{Step~3.7:}] Charlie announces publicly the authenticity of Alice
and Bob are simultaneously satisfied using an unjammable public channel. 
\end{description}

\subsection{Comparison of the existing and the proposed protocols\label{sec:Comparison}}

Criticisms of many of the existing protocols (including that of the
pioneering work of $\text{{\rm Cr�peau}}$ et al. \cite{CS_1995})
arise from the fact that they are based on the two-party secure multiparty
computational (two-party SMC) tasks which are not allowed in the domain
of non-relativistic quantum mechanics. To understand this particular
shortcoming of a class of protocols, we have to briefly state the
important results obtained by Lo in 1997 \cite{L97}, and Lo and Chau
in 1998 \cite{LC98}. These works essentially established that quantum
bit commitment, quantum remote coin tossing\footnote{Only a weak version of coin tossing can be implemented using quantum
resources.} and quantum oblivious transform (OT) cannot be performed with unconditional
security by using quantum means in two-party scenario. Here, it must
be noted that $\text{{\rm Cr�peau}}$ et al.'s pioneering protocol
\cite{CS_1995} of QIA was based on OT, but when it was published
then Lo-Chau results were not known. Interestingly, even after the
works of Lo and Chau, and many others in the same line \cite{BCS12,C07},
a large number of schemes for QIA have been proposed using computational
tasks which are apparently not allowed to be performed in an unconditionally
secure manner by the Lo Chau and similar results. For example, secure
quantum private comparison is not allowed in the two-party scenario,
but the protocols for QIA have been proposed using the schemes of
secure quantum private comparison. The question is whether Lo and
Chau results nullify the validity of all such schemes for QIA. The
answer is no. Here, we argue, why is it so. Further, the vulnerability
of the scheme of Mihara \cite{T.Mihara_2002} and how to circumvent
that is discussed above. In addition, it may be noted that there are
many theoretical proposals for QIA which requires quantum memory (for
example protocols described in Refs. \cite{ZW_1998,T.Mihara_2002,LB_2004,ZLG_2000,ZZZZ_2005,DXXN_2010,LLY_2006,WZT_2006,YTXZ_2013,TJ_2014,AHKB_2017,KHHYHM_2018,WZGZ_2019}).
As quantum memory in the true sense is not yet available, these schemes
of QIA are not good candidates for practical realization at the moment.
Of course, one can circumvent the need of quantum memory to some extent
using delay. However, that's a restricted choice. Here it would be
apt to note that even an effort to implement Protocol 3 proposed here
would face this technological hindrance. Another problem associated
with the implementation of QIA over long distance is the unavailability
of quantum repeaters. In some schemes, like Zhou et al. scheme for
QIA \cite{ZZZZ_2005} entanglement swapping is used to address this
issue. However, in such an approach the device in the middle used
for entanglement swapping needs to be trusted, which carries with
it a security concern. 

\section{Security analysis\textcolor{blue}{{} \label{sec:Security-analysis}}}

Here, we analyze the security of the proposed protocols against some
of the commonly discussed attacks by an outsider (Eve). Insider attacks
are not relevant in the context of identity authentication and consequently
they are not discussed here. To begin with, we will discuss impersonation
attack and discuss the security of all the protocols proposed above
(against this attack) in a sequential manner.

\subsection{Impersonation attack}

In this kind of attacks, an outsider Eve may try to impersonate the
legal user Alice (or Bob) and pass the authentication process. In
what follows, we will show that the proposed protocols are not vulnerable
under such an attack of Eve. 

\subsubsection{Security analysis on Protocol 1 against impersonation attack by Eve}

Let us consider the situation in which Eve is trying to impersonate
Alice. As per the protocol, $n$ authentication particles are to be
inserted in the $n$-particle decoy sequence. The probability of Eve
to correctly guess the correct authentication state is $\frac{1}{2}$.
Further, Eve has to correctly guess the position of decoy particle
too, and probability of that is also $\frac{1}{2}$. Hence the probability
of Eve to impersonate Alice is $(\frac{1}{4})^{n}$. So, the probability
of detecting Eve's presence is $P_{1}=1-(\frac{1}{4})^{n}$\textcolor{red}{{}
}and for large value of $n$, it ($P_{1})$ tends to 1 (see Fig.(\ref{fig:3 The-relationship-between})).
\textcolor{black}{In the case of Eve to impersonate Bob, the situation
will be similar to that for impersonating Alice.}

\subsubsection{Security analysis on Protocol 2 against impersonation attack by Eve}

Consider that Eve prepares a $2n$ random classical sequence and to
impersonate Alice, she computes $k_{2i-1}\oplus k_{2i}$. Here, probability
to get the same result as that of Alice would be $\frac{1}{2}$. Additionally,
the probability of Eve to prepare the correct state and send them
to Bob is $\frac{1}{2}$. So, the probability of detecting of Eve's
presence would be $P_{2}=1-(\frac{1}{4})^{n}$. Thus, against this
attack Protocol 2 will be as secure as Protocol 1 is.

\subsubsection{Security analysis of Protocol 3 against impersonation attack by Eve }

\begin{figure}[h]
\centering{}\includegraphics[scale=0.6]{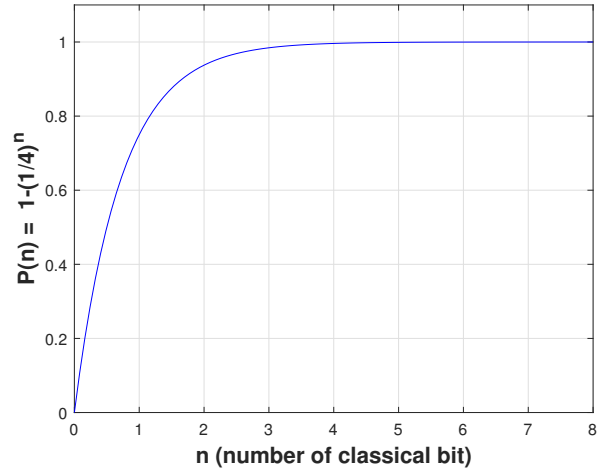}\caption{\label{fig:3 The-relationship-between}(Color online) The relationship
between the probability of detecting of Eve's presence $P(n)$ and
number of using classical bit $n$. }
\end{figure}

Without any loss of generality, we may consider the situation described
in Section \ref{subsec:Basic-idea}. Eve (impersonating Alice) chooses
$|\phi^{+}\rangle_{12}$ and sends the particle (particle 2) sequence
$S_{e2}$ to Charlie. Charlie follows the protocol and sends back
the sequence $S_{2e}$ ($S_{4}$) to Bob (Eve). Eve performs all the
steps of protocol as expected from real Alice and for this case will
apply Pauli operation $I$ on the Particle $1$. Bob and Eve perform
the Bell measurements and Table (\ref{tab:2}) shows all the possible
measurement scenarios that reveals the presence of Eve. As the correct
key $K_{AB}$ is not known to Eve,\textcolor{blue}{{} }so the probability
of Eve choosing the correct Bell state is $\frac{1}{4}$ and Eve has
to successfully guess all the $n$ Bell states in order to escape
from the security check. So, the probability of successful impersonation
attack by Eve is $(\frac{1}{4})^{n}$. If $n$ is very large, the
probability of the success of impersonation attack will be close to
zero. Thus, the probability $P_{3}$ to detect the presence of Eve
is $1-(\frac{1}{4})^{n}$. For large $n$ value the probability $P_{3}$
will be approximately 1 and impersonation attack can be identified.
It may be noted that $P_{1}=P_{2}=P_{3}=P(n)=1-(\frac{1}{4})^{n}$,
the relationship between $P(n)$ and $n$ is shown in\textcolor{blue}{{}
}Fig.(\ref{fig:3 The-relationship-between}) which clearly shows that
the minimum 6, 6 and 10 classical bits information is needed as pre-shared
key to detect the Eve's presence for Protocol 1, Protocol 3 and Protocol
2, respectively.

\begin{table}[h]
\begin{centering}
\caption{\label{tab:2}The possible measured results}
\par\end{centering}
\centering{}%
\begin{tabular*}{16.5cm}{@{\extracolsep{\fill}}ccc}
\toprule 
Eve and Bob's possible combined result & Charlie's result & Additional modulo 2\tabularnewline
\midrule
$|\psi^{+}\rangle_{14}\otimes|\psi^{-}\rangle_{23}$ & $|0\rangle_{5}$ & $10\otimes11=01$\tabularnewline
$\vert\psi^{-}\rangle_{14}\otimes|\psi^{+}\rangle_{23}$ & $|0\rangle_{5}$ & $11\otimes10=01$\tabularnewline
$\vert\phi^{+}\rangle_{14}\otimes|\phi^{-}\rangle_{23}$ & $|0\rangle_{5}$ & $00\otimes01=01$\tabularnewline
$\vert\phi^{-}\rangle_{14}\otimes|\phi^{+}\rangle_{23}$ & $|0\rangle_{5}$ & $01\otimes00=01$\tabularnewline
$\vert\psi^{+}\rangle_{14}\otimes|\phi^{-}\rangle_{23}$ & $|1\rangle_{5}$ & $10\otimes01=11$\tabularnewline
$\vert\psi^{-}\rangle_{14}\otimes|\phi^{+}\rangle_{23}$ & $|1\rangle_{5}$ & $11\otimes00=11$\tabularnewline
$\vert\phi^{+}\rangle_{14}\otimes|\psi^{-}\rangle_{23}$ & $|1\rangle_{5}$ & $00\otimes11=11$\tabularnewline
$\vert\phi^{-}\rangle_{14}\otimes|\psi^{+}\rangle_{23}$ & $|1\rangle_{5}$ & $01\otimes10=11$\tabularnewline
\bottomrule
\end{tabular*}
\end{table}

\subsection{Measurement and resend attack\label{subsec:Measurement-and-resend}}

\subsubsection{Security analysis on Protocol 1 against measure-resend attack\label{subsec:2.2.1}}

In this attack strategy, Eve will measure the sequence of qubits traveling
from Alice to Bob, and send new states to Bob depending upon the measurement
result. For Eve to impersonate Alice, it is advisable for Eve to separate
decoy qubits from the authentication qubits, but Eve has no information
about the position of the decoy states. Consequently, in her effort
to implement measure-resend attack, she eventually attacks the information
qubits, too and that leads to detectable traces of her presence. To
visualize this, let us assume that Eve attacks all the qubits traveling
through the channel (the analysis and conclusion will be similar even
if she attacks a fraction $f$ of the traveling qubits). Here, Eve
measures all the qubits of the sequence either in $\{|0\rangle,|1\rangle\}$
basis or in $\{|+\rangle,|-\rangle\}$ basis randomly (i.e., the probability
that a particular basis is used to measure a specific qubit is $\frac{1}{2}$).
Before going further, let us calculate mutual information between
Alice and Bob for the case of authentication particles under the presence
of Eve. Here $P(B,A)$ is the joint probability for getting measurement
result $B$ by Bob when Alice prepares the state $A$ and $P(B)$
be the total probability of getting result in state $B$ from Bob's
measurement. 

\[
P(B,A)=\left\{ \begin{array}{cc}
\frac{1}{8} & {\rm when}\,B\,{\rm and}A\,{\rm are\,meassured\,in\,the\,same\,basis\,but\,different\,outcomes\,are\,obtained}\\
\frac{3}{8} & {\rm when}\,B\,{\rm and}A\,{\rm are\,meassured\,in\,the\,same\,basis\,and\,the\,same\,outcomes\,are\,obtained}\\
0 & {\rm when}\,B\,{\rm and}A\,{\rm are\,meassured\,in\,the\,different\,basis\,}
\end{array}\,\right.
\]
and 
\[
P(B=|0\rangle)=P(B=|-\rangle)=\frac{3}{8}
\]

\[
P(B=|1\rangle)=P(B=|+\rangle)=\frac{1}{8}
\]
where $A\in\left\{ |0\rangle,|-\rangle\right\} $ and $B\in\left\{ |0\rangle,|1\rangle,|+\rangle,|-\rangle\right\} .$

The conditional entropy\footnote{Conditional probability, $P(B=|i\rangle|A=|j\rangle)=\frac{P(B=|i\rangle,A=|j\rangle)}{P(A=|j\rangle)}$
and conditional entropy, $H(B|A)=-\sum_{j}P(A=|j\rangle)\sum_{i}P(B=|i\rangle|A=|j\rangle)\log_{2}P(B=|i\rangle|A=|j\rangle)$.} is $H(B|A)=\left[2\times\frac{1}{2}\left(-\frac{3}{4}\log_{2}\frac{3}{4}-\frac{1}{4}\log_{2}\frac{1}{4}\right)\right]=0.811278$
bit, $H(A)=1$ bit and $H(B)=2\times\left(-\frac{3}{8}\log_{2}\frac{3}{8}-\frac{1}{8}\log_{2}\frac{1}{8}\right)=1.811278$
bit. The mutual information between Alice and Bob is $I(A:B)=H(B)-H(B|A)=1.0$
bit\footnote{The mutual information between two parties is bounded by the Shannon
entropy of probability distributions for a single particle, i.e.,
$0\leq I(A:B)\leq min[H(A),H(B)]$ as $H(A,B):=[max[H(A),H(B)],H(A)+H(B)],$
and $I(A:B):=H(A)+H(B)-[max[H(A),H(B)],H(A)+H(B)]],$ so $I(A:B):=[0,min[H(A),H(B)]]$. }. 

In a similar way, the probabilities for Alice and Eve are

\[
P(E,A)=\left\{ \begin{array}{cc}
0 & {\rm when}\,E\,{\rm and}A\,{\rm are\,meassured\,in\,the\,same\,basis\,but\,different\,outcomes\,are\,obtained}\\
\frac{1}{4} & {\rm when}\,E\,{\rm and}A\,{\rm are\,meassured\,in\,the\,same\,basis\,and\,the\,same\,outcomes\,are\,obtained}\\
\frac{1}{8} & {\rm when}\,E\,{\rm and}A\,{\rm are\,meassured\,in\,the\,different\,basis\,}
\end{array}\,\right.
\]
and 
\[
P(E=|0\rangle)=P(E=|-\rangle)=\frac{3}{8},
\]

\[
P(E=|1\rangle)=P(E=|+\rangle)=\frac{1}{8},
\]
where $A\in\left\{ |0\rangle,|-\rangle\right\} $and $E\in\left\{ |0\rangle,|1\rangle,|+\rangle,|-\rangle\right\} $.

The conditional entropy is $H(E|A)=\left[2\times\frac{1}{2}(-\frac{1}{2}\log_{2}\frac{1}{2}-\frac{1}{4}\log_{2}\frac{1}{4}-\frac{1}{4}\log_{2}\frac{1}{4}\right]=1.5$
bit and $H(E)=2\times\left(-\frac{3}{8}\log_{2}\frac{3}{8}-\frac{1}{8}\log_{2}\frac{1}{8}\right)=1.811278$.
The mutual information between Eve and Alice is $I(A:E)=H(E)-H(E|A)=0.311278$
bit. Further, we know that the mutual information between Alice and
Eve $I(A:E)$ is restricted by Holevo bound \cite{H_73} which for
this case is 0.600876. We can see that $I(A:E)\le0.600876$ and $I(A:B)>I(A:E)$,
which implies that Eve's information is considerably lesser than that
of Bob. 

To impersonate Bob, Eve has to attack the enlarged particle sequence
$K_{B}$ which is returned by Bob. This scenario is the same as that
required for impersonating Alice.

\subsubsection{Security analysis on Protocol 2 against measure-resend attack}

Eve performs a random measurement on particles sent by Alice and gets
some outcomes. Similar to Protocol 1, we can compute the mutual information
between Alice and Bob as 

\textbf{
\[
P(B,A)=\frac{1}{16}+\frac{1}{8}\delta_{A,B}\,{\rm and}\,P(B=|0\rangle)=P(B=|+\rangle)=(B=|1\rangle)=P(B=|-\rangle)=\frac{1}{4},
\]
$\delta_{A,B}=1(0)$ }when the measurement outcomes of Alice and Bob
are the same (different) using the same basis.\textbf{ }

The conditional entropy will be $H(B|A)=\frac{1}{4}\left[4\times\left(-\frac{3}{4}\log_{2}\frac{3}{4}\right)+4\times\left(-\frac{1}{4}\log_{2}\frac{1}{4}\right)\right]=0.81127$
bit and $H(A)=H(B)=4\times\left[-\frac{1}{4}\log_{2}\frac{1}{4}\right]=2$
bits. The mutual information between Alice and Bob is $I(A:B)=H(B)-H(B|A)=1.18872$
bit which is more than one because they have already identical classical
information as pre-shared key.

For mutual information between Alice and Eve, 
\[
P(E,A)=\left\{ \begin{array}{cc}
0 & {\rm when}E\,{\rm and}A\,{\rm are\,meassured\,in\,the\,same\,basis\,but\,different\,outcomes\,are\,obtained}\\
\frac{1}{8} & {\rm when}E\,{\rm and}A\,{\rm are\,meassured\,in\,the\,same\,basis\,and\,the\,same\,outcomes\,are\,obtained}\\
\frac{1}{16} & {\rm when}E\,{\rm and}A\,{\rm are\,meassured\,in\,the\,different\,basis\,}
\end{array}\,\right.
\]
and
\[
P(E=|0\rangle)=P(E=|1\rangle)=P(E=|+\rangle)=P(E=|-\rangle)=\frac{1}{4}.
\]
So, conditional entropy is $H(E|A)=\frac{1}{4}\times[8\times(-\frac{1}{4}\log_{2}\frac{1}{4})-4\times(-\frac{1}{2}\log_{2}\frac{1}{2})]=1.5$
bit and $H(E)=4\times[-\frac{1}{4}\log_{2}\frac{1}{4}]=2$ bits. The
mutual information between Eve and Alice is $I(A:E)=H(E)-H(E|A)=0.5$
bit. Further, we know that the mutual information between Alice and
Eve, $I(A:E)$ is bound by Holevo quantity \cite{H_73} which is 1
in this case. We can see that $I(A:E)\le1$ and $I(A:B)>I(A:E),$
which satisfies the requirement for a secure communication without
disclosing meaningful information to Eve. Similar are the results
for the case when Eve tries to impersonate Bob.

\subsubsection{Security analysis on Protocol 3 against measurement and resend attack}

In this protocol, Alice and Bob prepares the Bell states depending
upon their pre-shared secret key. Alice and Bob both do not dispatch
their quantum states as a whole. So Eve has to attack only one particle
at a time. Without loss of generality, we consider that Eve will attack
Particle $2$ (Particle $4$) when Alice sends it to Charlie (Charlie
sends it to Alice). Eve may extract the maximum information from the
quantum channel which is restricted by Holevo bound or Holevo quantity
\cite{H_73}

\begin{equation}
\chi(\rho)=S(\rho)-\sum_{i}p_{i}S(\rho_{i}),\label{eq:8}
\end{equation}
Where $S(\rho)=-Tr(\rho\log_{2}\rho)$ is the von Neumann entropy,
$\rho_{i}$ is a component in the mixed state $\rho$ with probability
$p_{i}$. We have already discussed that Eve will attack on Particle
$2$ ($4$), for this situation (\ref{eq:8}) can be rewritten as:

\begin{equation}
\chi(\rho_{2})=S(\rho_{2})-\sum_{i}p_{i}S(\rho_{2_{i}})\label{eq:9}
\end{equation}
where $\rho_{2}$ and $\rho_{2_{i}}$ are the reduced density matrix
of $\rho$ and $\rho_{i}$ respectively after partial trace over 1,
4, 3, 5 particles. From (\ref{eq:5}), we can write $\rho=[|\Psi\rangle\langle\Psi|]_{14235}$.
The reduced density matrix for Particle $2$ (4) always has the form, 

\[
\rho_{2}=Tr_{1435}\left(|\Psi\rangle\langle\Psi|\right)_{14235}=\frac{1}{2}I
\]
and $\rho_{2_{i}}$will be the corresponding reduced density matrix.
The probability of choosing another combination of authentication
key by Alice and Bob is $\frac{1}{4}$ i.e., here $p_{i}=\frac{1}{4}$.
Here, also $\rho_{2_{i}}=\frac{1}{2}I.$ Substituting $\rho_{2}$
and $\rho_{2_{i}}$ into (\ref{eq:9}) gives, $\chi(\rho_{2})=0$.
Similar result can be obtained for Particle $4$. This result implies
Eve cannot get any information in direct measurement attack on Particle
$2$ (and $4$).

\subsection{Impersonated fraudulent attack}

\subsubsection{Security analysis on Protocol 1 against impersonated fraudulent attack
strategy by forging new qubits}

Here, Eve tries to impersonate Alice by introducing a unitary operation
$U_{E}$ to correlate Alice's qubit with an ancilla qubit. After application
of unitary operator, the quantum state can be written as

\[
U_{E}|0\chi\rangle_{Ae}=|\psi\rangle_{0}=a_{0}|00\rangle+b_{0}|01\rangle+c_{0}|0+\rangle+d_{0}|0-\rangle,
\]

\[
U_{E}|1\chi\rangle_{Ae}=|\psi\rangle_{1}=a_{1}|10\rangle+b_{1}|11\rangle+c_{1}|1+\rangle+d_{1}|1-\rangle,
\]

\[
U_{E}|+\chi\rangle_{Ae}=|\psi\rangle_{+}=a_{+}|+0\rangle+b_{+}|+1\rangle+c_{+}|++\rangle+d_{+}|+-\rangle,
\]

\[
U_{E}|-\chi\rangle_{Ae}=|\psi\rangle_{-}=a_{-}|-0\rangle+b_{-}|-1\rangle+c_{-}|-+\rangle+d_{-}|--\rangle.
\]
So, we obtain the entire state as,\textcolor{blue}{{} }

\begin{equation}
|\rho\rangle=\frac{1}{4}\left(|\psi\rangle_{00}\langle\psi|+|\psi\rangle_{11}\langle\psi|+|\psi\rangle_{++}\langle\psi|+|\psi\rangle_{--}\langle\psi|\right).\label{eq:10}
\end{equation}

The probability of getting the correct measurement result for each
state of authentication sequence as well as the decoy sequence by
Bob is $\frac{1}{2}$. Suppose, we take the correct state as $|0\rangle$,
then the probability of detecting Eve is $P_{0}=\frac{1}{2}$. Similarly,
we obtain $P_{1}=P_{+}=P_{-}=\frac{1}{2}$. The detection possibility
for each qubit is $P_{d}=\frac{1}{4}(P_{0}+P_{1}+P_{+}+P_{-})=\frac{1}{2}$.
According to the Simmons theory \cite{S_88}, the protocol is unconditionally
secure under impersonated fraudulent attack using (\ref{eq:10}).

\subsubsection{Security analysis on Protocol 2 against impersonated fraudulent attack
strategy by forging new qubits}

In this protocol, Alice only uses the authentication state in the
computational and Hadamard basis with equal probability and the expected
measurement result by Bob also equally probable. So, the attack situation
is similar to the Protocol 1. In this context, Protocol 2 is also
secure under the impersonated fraudulent attack.

\subsubsection{Security analysis on Protocol 3 against impersonated fraudulent attack
strategy by forging new qubits}

To impersonate Alice, an optimal strategy by Eve is to operate the
traveling particle (particle 2) sent by Alice with her ancillary state.
After operating the general operation with traveling particle, the
state can be denoted as follows,

\begin{equation}
U_{E}|1\chi\rangle_{2e}=\left(a_{0}|10\rangle+b_{0}|11\rangle+c_{0}|00\rangle+d_{0}|01\rangle\right)_{2e},
\end{equation}

\begin{equation}
U_{E}|0\chi\rangle_{2e}=\left(a_{1}|10\rangle+b_{1}|11\rangle+c_{1}|00\rangle+d_{1}|01\rangle\right)_{2e},
\end{equation}

Here, $|\chi\rangle_{e}$ is the ancillary state which is created
by Eve, the subscript $e$ refers to state prepared by Eve and $2$
refers the traveling particle of Alice. For normalization $|a_{0}^{2}|+|b_{0}^{2}|+|c_{0}^{2}|+|d_{0}^{2}|=|a_{1}^{2}|+|b_{1}^{2}|+|c_{1}^{2}|+|d_{1}^{2}|=1$.

Eve's operation creates the following state,

\begin{equation}
|\Psi^{\prime}\rangle_{12e}=\frac{1}{\sqrt{2}}\left(a_{0}|010\rangle+b_{0}|011\rangle+c_{0}|000\rangle+d_{0}|001\rangle+a_{1}|110\rangle+b_{1}|111\rangle+c_{1}|100\rangle+d_{1}|101\rangle\right)\label{eq:13}
\end{equation}

For simplicity, Eve employs a general operation on the traveling particle
and keep her own qubit with herself and sends the Alice's particle
to Charlie. Charlie does the necessary operation and returns this
particle according to protocol. Here we aim to calculate the final
composite state after consideration of Eve's attack along with\textcolor{blue}{{}
}the scenario described in Section \ref{subsec:Basic-idea} as follows,

\begin{equation}
\begin{array}{lcl}
|\Psi^{\prime\prime}\rangle_{12e345} & = & \frac{1}{2\sqrt{2}}[a_{0}|110110\rangle-a_{0}|100111\rangle-a_{0}|110000\rangle+a_{0}|100001\rangle\\
 & + & b_{0}|111110\rangle-b_{0}|101111\rangle-b_{0}|111000\rangle+b_{0}|101001\rangle\\
 & - & c_{0}|100110\rangle-c_{0}|110111\rangle-c_{0}|100000\rangle+c_{0}|110001\rangle\\
 & + & d_{0}|101110\rangle-d_{0}|111111\rangle-d_{0}|101000\rangle+d_{0}|111001\rangle\\
 & - & a_{1}|010110\rangle+a_{1}|000111\rangle+a_{1}|010000\rangle+a_{1}|000001\rangle\\
 & - & b_{1}|011110\rangle+b_{1}|001111\rangle+b_{1}|011000\rangle-b_{1}|001001\rangle\\
 & - & c_{1}|000110\rangle+c_{1}|010111\rangle+c_{1}|000000\rangle-c_{1}|010001\rangle\\
 & - & d_{1}|001110\rangle+d_{1}|011111\rangle+d_{1}|001000\rangle-d_{1}|011001\rangle]
\end{array}\label{eq:CNOT operation}
\end{equation}

After receiving the particle 4 from Charlie, Eve and Bob perform the
Bell measurement on their particles that they have and sends the classical
sequence of measurement result to Charlie. By applying the authentication
condition, we can easily find the success probability and the failure
probability of Eve to pass the authentication is $\frac{1}{4}(|b_{0}|^{2}+|c_{0}|^{2})+\frac{1}{8}$
and $\frac{1}{4}(|a_{0}|^{2}+|d_{0}|^{2}+1)+\frac{3}{8}$ respectively.
Suppose that Eve prepares the ancillary state such that the value
of $|b_{0}|^{2}+|c_{0}|^{2}$ will be maximum and the value of $|a_{0}|^{2}+|d_{0}|^{2}$
will be minimum, then she can deceives maximally as an authenticated
member with probability $P_{3}^{2}=\frac{3}{8}$. For rest of the
cases, $P_{3}^{2}\leq\frac{3}{8}$. From this discussion, it can be
concluded that the protocol is secure under some acceptable error
limit for the best case scenario of collective attack by Eve.

\section{Conclusions\label{sec:Conclusions}}

We have reviewed the existing protocols for QIA with an intention
to understand the intrinsic symmetry among them. The analysis has
revealed the symmetry among various protocols and that has helped
us in classifying the existing protocols for QIA and to identify various
simple strategies which may be used to transform the existing protocols
for different quantum computation and communication tasks into new
schemes for QIA. To establish this point, a few new protocols for
QIA are also proposed and established to be secure against a set of
potential attacks. Interestingly, among the protocols discussed above
(including the existing and the newly proposed protocols) all are
not really implementable with the present technology. For example,
there are several protocols where one of the users need to keep a
photon (for example, consider ping-pong type scheme for QIA described
in \cite{ZZZX_2006} or the authentication schemes described in \cite{T.Mihara_2002,LB_2004,ZLG_2000,ZZZZ_2005,WZT_2006,TJ_2014,KHHYHM_2018,WZGZ_2019})
with him/her until travel photon(s) is (are) returned to him/her.
Such protocols of QIA would require quantum memory which is not available
at this moment. In fact, most of the entangled state based QSDC type
protocols for QIA and entangled state based protocols which are based
on schemes for quantum private comparison and the schemes for QIA
involving a third party, Trent who keeps a photon with himself, will
face the same problem. Thus, at the moment, protocols for QIA which
does not require quantum memory should be preferred. There are many
proposals for building quantum memory and these schemes (schemes requiring
quantum memory) may be useful in future, but direct teleportation
based schemes for QIA (e.g., schemes proposed in \cite{ZZZZ_2005,TJ_2014})
are not expected to find many applications, even in future. This is
so because teleportation can be directly used for secure quantum communication
only when the quantum channel is noise free. Further, any scheme that
requires shared entanglement (e.g., schemes described in \cite{CS_01,CSPF_02,ZLG_2000,T.Mihara_2002,LB_2004,ZZZZ_2005})
may require entanglement purification or concentration (as the shared
entanglement will disintegrate due to decoherence) which in turn would
require unwanted and unsafe interaction between Alice and Bob. Despite
these limitations and technological issues, the domain of QIA related
research is growing fast because the claimed unconditional security
of quantum cryptography schemes is also based on the security of identity
authentication scheme used. 

This review would remain incomplete, unless we mention that a large
number of post-quantum authentication schemes \cite{WZW+21,MGB+20}
have also been proposed in the recent past. We have not discussed
those schemes as they are classical in nature and conditionally secure.
A very strong assumption behind these schemes is that a quantum computer
will not be able to efficiently solve problems outside bounded-error
quantum polynomial time (BQP) complexity class. There is no such proof
and this assumption is technically equivalent to presuppose that if
we don't know any efficient algorithm for a given computational task
at the moment, no one will be able to construct that in future, too. 

\subsection*{Acknowledgment: }

The authors thank DRDO India for the support provided through the
project number ANURAG/MMG\\*/ CARS/2018-19/071. The authors also thank
Dr. Kishore Thapliyal and Dr. Sandeep Mishra for their interest and
technical feedback on this work.

\bibliographystyle{unsrt}
\bibliography{authentication-review}

\end{document}